\documentstyle{elsart}

\makeatletter
\def\Tr{\mathop{\operator@font Tr}\nolimits}
\def\Re{\mathop{\operator@font Re}\nolimits}
\def\Im{\mathop{\operator@font Im}\nolimits}
\def\goto{\mathop{\;\longrightarrow\;}}
\def\I{{\mathrm{i}}}
\def\d{{\mathrm{d}}}
\def\princint{\ \raise 0.55 ex\hbox to 0.6em{\hrulefill}%
\kern-0.88em\int}
\def\casefr#1#2{{\textstyle{#1\over#2}}}
\def\vereq#1#2{\lower3pt\vbox{\baselineskip1.5pt \lineskip1.5pt
\ialign{$\m@th#1\hfill##\hfil$\crcr#2\crcr\sim\crcr}}}
\def\lesssim{\mathrel{\mathpalette\vereq<}}
\def\gtrsim{\mathrel{\mathpalette\vereq>}}
\@addtoreset{equation}{section}%
\def\@journal{Physics Reports}
\def\@date{July 1996}
\makeatother

\begin{document}
\begin{frontmatter}


\title{The large-$N$ expansion of unitary-matrix models}

\author{Paolo Rossi, Massimo Campostrini, and Ettore Vicari}
\address{Dipartimento di Fisica dell'Universit\`a and I.N.F.N.,
I-56126 Pisa, Italy}

\begin{abstract}
The general features of the $1/N$ expansion in statistical mechanics
and quantum field theory are briefly reviewed both from the
theoretical and from the phenomenological point of view as an
introduction to a more detailed analysis of the large-$N$ properties
of spin and gauge models possessing the symmetry group ${\rm SU}(N)
\times {\rm SU}(N)$.

An extensive discussion of the known properties of the single-link
integral (equivalent to YM$_2$ and one-dimensional chiral models)
includes finite-$N$ results, the external field solution, properties
of the determinant, and the double scaling limit. 

Two major classes of solvable generalizations are introduced:
one-dimensional closed chiral chains and models defined on a $d-1$
dimensional simplex.  In both cases large-$N$ solutions are presented
with emphasis on their double scaling properties.

The available techniques and results concerning unitary-matrix models
that correspond to asymptotically free quantum field theories
(two-dimensional chiral models and four-dimensional QCD) are
discussed, including strong-coupling methods, reduced formulations,
and the Monte Carlo approach.
\end{abstract}


\end{frontmatter}




\newpage
\begingroup
\parskip 0pt 
\tableofcontents
\endgroup


\section{Introduction}
\label{introduction}

\subsection{General motivation for the $1/N$ expansion}
\label{intro-general}

The approach to quantum field theory and statistical mechanics based
on the identification of the large-$N$ limit and the perturbative
expansion in powers of $1/N$, where $N$ is a quantity related to the
number of field components, is by now almost thirty years old.  It
goes back to the original work by Stanley \cite{Stanley-II} on the
large-$N$ limit of spin systems with ${\rm O}(N)$ symmetry, soon
followed by Wilson's suggestion that the $1/N$ expansion may be a
valuable alternative in the context of renormalization-group
evaluation of critical exponents, and by 't Hooft's extension
\cite{THooft-planar} to gauge theories and, more generally, to fields
belonging to the adjoint representation of ${\rm SU}(N)$ groups.  More
recently, the large-$N$ limit of random-matrix models was put into a
deep correspondence with the theory of random surfaces, and therefore
it became relevant to the domain of quantum gravity.

In order to understand why the $1/N$ expansion should be viewed as a
fundamental tool in the study of quantum and statistical field theory,
it is worth emphasizing a number of relevant features:

1) $N$ is an intrinsically dimensionless parameter, representing a
dependence whose origin is basically group-theoretical, and leading to
well-defined field representations for all integer values, hence it is
not subject to any kind of renormalization;

2) $N$ does not depend on any physical scale of the theory, hence we
may expect that physical quantities should not show any critical
dependence on $N$ (with the possible exception of finite-$N$ scaling
effects in the double-scaling limit);

3) the large-$N$ limit is a thermodynamical limit, in which we observe
the suppression of fluctuations in the space of internal degrees of
freedom; hence we may expect notable simplifications in the algebraic
and analytical properties of the model, and even explicit
integrability in many instances.

Since integrability does not necessarily imply triviality, the
large-$N$ solution to a model may be a starting point for finite-$N$
computations, because it shares with interesting finite values of $N$
many physical properties.  (This is typically not the case for the
standard free-field solution which forms the starting point for the
usual perturbative expansions.)  Moreover, for reasons which are
clearly, if not obviously, related to the three points above, the
physical variables which are naturally employed to parameterize
large-$N$ results and $1/N$ expansions are usually more directly
related to the observables of the models than the fields appearing in
the original local Lagrangian formulation.

More reasons for a deep interest in the study of the large-$N$
expansion will emerge from the detailed discussion we shall present in
the rest of this introductory section.  We must however anticipate
that many interesting review papers have been devoted to specific
issues in the context of the large-$N$ limit, starting from Coleman's
lectures \cite{Coleman-erice}, going through Yaffe's review on the
reinterpretation of the large-$N$ limit as classical mechanics
\cite{Yaffe}, Migdal's review on loop equations
\cite{Migdal-equations}, and Das' review on reduced models
\cite{Das-review}, down to Polyakov's notes \cite{Polyakov-book} and
to the recent large commented collection of original papers by Brezin
and Wadia \cite{Brezin-Wadia}, not to mention Sakita's booklet
\cite{Sakita-book} and Ma's contributions
\cite{Ma-introduction,Ma-largeN}.  Moreover, the $1/N$ expansion of
two-dimensional spin models has been reviewed by two of the present
authors a few years ago \cite{Campostrini-Rossi-review}.  As a
consequence, we decided to devote only a bird's eye overview to the
general issues, without pretension of offering a self-contained
presentation of all the many conceptual and technical developments
that have appeared in an enormous and ever-growing literature; we even
dismissed the purpose of offering a complete reference list grouped by
arguments, because the task appeared to be beyond our forces.

We preferred to focus on a subset of all large-$N$ topics, which has
never been completely and systematically reviewed: the issue of
unitary-matrix models.  Our self-imposed limitation should not appear
too restrictive, when considering that it still involves such topics
as ${\rm U}(N) \times {\rm U}(N)$ principal chiral models, virtually
all that concerns large-$N$ lattice gauge theories, and an important
subset of random-matrix models with their double-scaling limit
properties, related to two-dimensional conformal field theory.

The present paper is organized on a logical basis, which will neither
necessarily respect the sequence of chronological developments, nor
it will keep the same emphasis that was devoted by the authors of the
original papers to the discussion of the different issues.

Sect.\ \ref{unitary-matrices} is devoted to a presentation of the
general and common properties of unitary-matrix models, and to an
analysis of the different approaches to their large-$N$ solution that
have been discussed in the literature.

Sect.\ \ref{single-link} is a long and quite detailed discussion of the
most elementary of all unitary-matrix systems.  Since all essential
features of unitary-matrix models seem to emerge already in the
simplest example, we thought it worthwhile to make this discussion as
complete and as illuminating as possible.  

Sect.\ \ref{2d-YM} is an application of results obtained by studying
the single-link problem, which exploits the equivalence of this model
with lattice YM$_2$ and principal chiral models in one dimension.

Sect.\ \ref{chiral-chains} is devoted to a class of reasonably simple
systems, whose physical interpretation is that of closed chiral chains
as well as of gauge theories on polyhedra.

Sect.\ \ref{simplicial-chiral} presents another class of integrable
systems, corresponding to chiral models defined on a $d$-dimensional
simplex, whose properties are relevant both in the discussion of the
strong-coupling phase of more general unitary-matrix models and in the
context of random-matrix models.

Sect.\ \ref{principal-chiral} deals with the physically more
interesting applications of unitary-matrix models: two-dimensional
principal chiral models and four-dimensional lattice gauge theories,
sharing the properties of asymptotic freedom and ``confinement'' of
the Lagrangian degrees of freedom.  Special issues, like numerical
results and reduced models, are considered.

\subsection{Large $N$ as a thermodynamical limit: factorization} 
\label{intro-factorization}

As we already mentioned briefly in the introduction, one of the
peculiar features of the large-$N$ limit is the occurrence of notable
simplifications, that become apparent at the level of the quantum
equations of motion, and tend to increase the degree of integrability
of the systems.  These simplifications are usually related to a
significant reduction of the number of algebraically-independent
correlation functions, which in turn is originated by the property of
factorization.

This property is usually stated as follows: connected Green's
functions of quantities that are invariant under the full symmetry
group of the system are suppressed with respect to the corresponding
disconnected parts by powers of $1/N$.  Hence when $N\to\infty$ one
may replace expectation values of products of invariant quantities
with products of expectation values.

One must however be careful, since factorization is not a property
shared by all invariant operators without further qualifications.  In
particular, experience shows that operators associated with very high
rank representations of the symmetry group, when the rank is $O(N)$,
do not possess the factorization property.  A very precise
characterization has been given by Yaffe \cite{Yaffe}, who showed that
factorization is a property of ``classical'' operators, i.e., those
operators whose coherent state matrix elements have a finite
$N\to\infty$ limit.

It is quite interesting to investigate the physical origin of
factorization.  The property
\begin{equation}
\lim_{N\to\infty} \left<AB\right> = \left<A\right>\left<B\right>
\end{equation}
implies in particular that
\begin{equation}
\lim_{N\to\infty} \langle A^2\rangle = \left<A\right>^2,
\end{equation}
i.e., the vacuum state of the model, seen as a statistical ensemble,
seems to possess no fluctuations.  To be more precise, all the field
configurations that correspond to a nonvanishing vacuum wavefunction
can be related to each other by a symmetry transformation.  This
residual infinite degeneracy of the vacuum configurations makes the
difference between the large-$N$ limit and a strictly classical limit
$\hbar\to0$, and allows the possibility of violations of factorization
when infinite products of operators are considered; this is in a sense
the case with representations whose rank is $O(N)$.

More properly, we may view large $N$ as a thermodynamical limit
\cite{Haan}, since the number of degrees of freedom goes to
infinity faster than any other physical parameter, and as a
consequence the ``macroscopic'' properties of the system, i.e., the
invariant expectation values, are fixed in spite of the great number
of different ``microscopic'' realizations.  This realization does not
rule out the possibility of searching for the so-called ``master
field'', that is a representative of the equivalence class of the
field configurations corresponding to the large-$N$ vacuum, such that
all invariant expectation values of the factorized operators can be
obtained by direct substitution of the master field value into the
definition of the operators themselves \cite{Coleman-erice}.

There has been an upsurge of interest on master fields in recent years
\cite{Gopakumar-Gross,Douglas-master}, triggered by new results in
non-commutative probability theory applied to the stochastic master
field introduced in Ref.\ \cite{Greensite-Halpern-master}.

\subsection{$1/N$ expansion of vector models in
statistical mechanics and quantum field theory} 
\label{intro-vector}

The first and most successful application of the approach based on the
large-$N$ limit and the $1/N$ expansion to field theories is the
analysis of vector models enjoying ${\rm O}(N)$ or ${\rm SU}(N)$
symmetry.  Actually, ``vector models'' is a nickname for a wide class
of different field theories, characterized by bosonic or fermionic
Lagrangian degrees of freedom lying in the fundamental representation
of the symmetry group (cfr.\ Ref.\ \cite{Campostrini-Rossi-review} and
references therein).

A quite general feature of these models is the possibility of
expressing all self-interactions of the fundamental degrees of freedom
by the introduction of a Lagrange multiplier field, a boson and a
singlet of the symmetry group, properly coupled to the Lagrangian
fields, such that the resulting effective Lagrangian is quadratic in
the $N$-component fields.  One may therefore formally perform the
Gaussian integration over these fields, obtaining a form of the
effective action which is nonlocal, but depends only on the singlet
multiplier, acting as a collective field; in this action $N$ appears
only as a parameter.

The considerations developed in Subs.\ \ref{intro-factorization} make
it apparent that all fluctuations of the singlet field must be
suppressed in the large-$N$ limit (no residual degeneracy is left in
the trivial representation).  As a consequence, solving the models in
this limit simply amounts to finding the singlet field configuration
minimizing the effective action.  The problem of nonlocality is easily
bypassed by the consideration that translation invariance of the
physical expectation values requires the action-minimizing field
configuration to be invariant in space-time; hence the saddle-point
equations of motion become coordinate-independent and all nonlocality
disappears. 

As one may easily argue from the above considerations, the large-$N$
solution of vector models describes some kind of Gaussian field
theory.  Nevertheless, this result is not as trivial as one might
imagine, since the free theory realization one is faced with usually
enjoys quite interesting properties, in comparison with the na{\"\i}ve
Lagrangian free fields.  Typical phenomena appearing in the large-$N$
limit are an extension of the symmetry and spontaneous mass
generation.  Moreover, when the fundamental fields possess some kind
of gauge symmetry, one may also observe dynamical generation of
propagating gauge degrees of freedom; this is the case with
two-dimensional ${\rm CP}^{N-1}$ models and their generalizations
\cite{Witten-CPN,DAdda-Luscher-DiVecchia-1/N}.

The existence of an explicit form of the effective action offers the
possibility of a systematic expansion in powers of $1/N$.  The
effective vertices of the theory turn out to be Feynman integrals over
a single loop of the free massive propagator of the fundamental field.
In two dimensions, where the physical properties of many vector models
are especially interesting (e.g., asymptotic freedom), these one-loop
integrals can all be computed analytically in the continuum version,
and even on the lattice many analytical results have been obtained.

The $1/N$ expansion is the starting point for a systematic computation
of critical exponents, which are nontrivial in the range $2<d<4$, for
the study of renormalizability of superficially nonrenormalizable
theories in the same dimensionality range, and for the computation of
physical amplitudes.  Notable is the case of the computation of
amplitude ratios, which are independent of the coupling in the scaling
region, and therefore are functions of $1/N$ alone; hopefully, their
$1/N$ expansion possesses a nonvanishing convergence radius.  The
$1/N$ expansion was also useful to explore the double-scaling limit
properties of vector models
\cite{Nishigaki-Yoneya,DiVecchia-Kato-Ohta,Damgaard-Heller}.

The properties of the large-$N$ limit and of the $1/N$ expansion of
continuum and lattice vector models were already reviewed by many
authors.  We therefore shall not discuss this topic further.  We only
want to stress that this kind of studies can be very instructive,
given the physical interest of vector models as realistic prototypes
of critical phenomena in two and three dimensions and as models for
dynamical Higgs mechanism in four dimensions.  Moreover, some of the
dynamical properties emerging mainly from the large-$N$ studies of
asymptotically free models (in two dimensions) may be used to mimic
some of the features of gauge theories in four dimensions; however,
at least one of the essential aspects of gauge theories, the presence
of matrix degrees of freedom (fields in the adjoint representation),
cannot be captured by any vector model.

\subsection{$1/N$ expansion of matrix models: planar diagrams}
\label{intro-planar-diagrams}

The first major result concerning the large-$N$ limit of matrix-valued
field theories was due to G. 't Hooft, who made the crucial
observation that, in the $1/N$ expansion of continuum gauge theories,
the set of Feynman diagrams contributing to any given order admits a
simple topological interpretation.  More precisely, by drawing the
${\rm U}(N)$ fundamental fields (``quarks'') as single lines and the
${\rm U}(N)$ adjoint fields (``gluons'') as double lines, each line
carrying one color index, a graph corresponding to a $n$th-order
contribution can be drawn on a genus $n$ surface (i.e., a surface
possessing $n$ ``holes'').  In particular, the zeroth-order
contribution, i.e., the large-$N$ limit, corresponds to the sum of all
planar diagrams.  The extension of this topological expansion to gauge
models enjoying ${\rm O}(N)$ and ${\rm Sp}(2N)$ symmetry has been
described by Cicuta \cite{Cicuta}.  Large-$N$ universality among ${\rm
O}(N)$, ${\rm U}(N)$, and ${\rm Sp}(2N)$ lattice gauge theories has
been discussed by Lovelace \cite{Lovelace-universality}.

This property has far-reaching consequences: it allows for
reinterpretations of gauge theories as effective string theories, and
it offers the possibility of establishing a connection between matrix
models and the theory of random surfaces, which will be exploited in
the study of the double-scaling limit.  

As a byproduct of this analysis, 't Hooft performed a summation of all
planar diagrams in two-dimensional continuum Yang-Mills theories, and
solved QCD$_2$ to leading nontrivial order in $1/N$,
finding the meson spectrum \cite{THooft-mesons,Callan-Coote-Gross}.

Momentum-space planarity has a coordinate-space counterpart in lattice
gauge theories.  It is actually possible to show that, within the
strong-coupling expansion approach, the planar diagrams surviving in
the large-$N$ limit can be identified with planar surfaces built up of
plaquettes by gluing them along half-bonds
\cite{Kazakov-pl,OBrien-Zuber-expansion,Kostov-sc}.  This construction
however leads quite far away from the simplest model of planar random
surfaces on the lattice originally proposed by Weingarten
\cite{Weingarten-pathological,Weingarten-nonplanar}, and hints at some
underlying structure that makes a trivial free-string interpretation
impossible.

\subsection{The physical interpretation: QCD phenomenology}
\label{intro-QCD}

The sum of the planar diagrams has not till now been performed in the
physically most interesting case of four-dimensional ${\rm SU}(N)$
gauge theories.  It is therefore strictly speaking impossible to make
statements about the relevance of the large-$N$ limit for the
description of the physically relevant case $N=3$.  However, it is
possible to extract from the large-$N$ analysis a number of
qualitative and semi-quantitative considerations leading to a very
appealing picture of the phenomenology predicted by the $1/N$
expansion of gauge theories.  These predictions can be improved
further by adopting Veneziano's form of the large-$N$ limit
\cite{Veneziano-unified}, in which not only the number of colors $N$
but also the number of flavors $N_f$ is set to infinity, while their
ratio $N/N_f$ is kept finite.  We shall not enter a detailed
discussion of large-$N$ QCD phenomenology, but it is certainly useful
to quote the relevant results.

\subsubsection{The large-$N$ property of mesons}

Mesons are stable and noninteracting; their decay amplitudes are
$O(N^{-1/2})$, and their scattering amplitudes are $O(N^{-1})$.

Meson masses are finite.

The number of mesons is infinite.

Exotics are absent and Zweig's rule holds.

\subsubsection{The large-$N$ property of glueballs}

Glueballs are stable and noninteracting, and they do not mix with
mesons; a vertex involving $k$ glueballs and $n$ mesons is
$O(N^{1-k-n/2})$.

The number of glueballs is infinite.

\subsubsection{The large-$N$ property of baryons}

A large-$N$ baryon is made out of $N$ quarks, and therefore it
possesses peculiar properties, similar of those of solitons
\cite{Witten-baryons}.

Baryon masses are $O(N)$.

The splitting of excited states is $O(1)$.

Baryons interact strongly with each other; typical vertices are
$O(N)$.

Baryons interact with mesons with $O(1)$ couplings.

\subsubsection{The $\eta'$ mass formula}

The spontaneous breaking of the ${\rm SU}(N_f)$ axial symmetry in QCD
gives rise to the appearance of a multiplet of light pseudoscalar
mesons.  This symmetry-breaking pattern was explicitly demonstrated in
the context of large-$N$ QCD by Coleman and Witten
\cite{Coleman-Witten}.  However, the singlet pseudoscalar is not
light, due to the anomaly of the ${\rm U}(1)$ axial current.  Since
the anomaly equation
\begin{equation}
\partial_\mu J_\mu^5 = {g^2 N_f \over 16\pi^2}
\Tr \widetilde F_{\mu\nu} F^{\mu\nu}
\end{equation}
has a vanishing right-hand side in the limit $N_c\to\infty$ with $N_f$
and $g^2N_c$ fixed (the standard large-$N$ limit of non-Abelian gauge
theories), the leading-order contribution to the mass of the $\eta'$
should be $O(1/N_c)$.  The proportionality constant should be related
to the symmetry-breaking term, which in turn is related to the
so-called topological susceptibility, i.e., the vacuum expectation
value of the square of the topological charge.  The resulting
relationship shows a rather satisfactory quantitative agreement with
experimental and numerical results \cite{Witten-algebra,%
DiVecchia-Veneziano,Rosenzweig-Schechter-Trahern,Nath-Arnowitt,%
Teper-topol-SU3}.

\subsection{The physical interpretation: two-dimensional quantum
    gravity} 
\label{intro-2dqg}

In the last ten years, a new interpretation of the $1/N$ expansion of
matrix models has been put forward.  Starting from the relationship
between the order of the expansion and the topology of two-dimensional
surfaces on which the corresponding diagrams can be drawn, several
authors \cite{Ambjorn-Durhuus-Frohlich,David-planar,David-random,%
Kazakov-bilocal,Kazakov-Kostov-Migdal} proposed that large-$N$ matrix
models could provide a representation of random lattice
two-dimensional surfaces, and in turn this should correspond to a
realization of two-dimensional quantum gravity.  These results were
found consistent with independent approaches, and proper modifications
of the matrix self-couplings could account for the incorporation of
matter.

The functional integrals over two-dimensional closed Riemann manifolds
can be replaced by the discrete sum over all (piecewise flat)
manifolds associated with triangulations.  It is then possible to
identify the resulting partition function with the vacuum energy
\begin{equation}
E_0 = - \log Z_N,
\end{equation}
obtained from a properly defined $N\times N$ matrix model, and the
topological expansion of two-dimensional quantum gravity is nothing
but the $1/N$ expansion of the matrix model.

The partition function of two-dimensional quantum gravity is expected
to possess well-defined scaling properties
\cite{Knizhnik-Polyakov-Zamolodchikov}.  These may be recovered in the
matrix model by performing the so-called ``double-scaling limit''
\cite{Brezin-Kazakov,Douglas-Shenker,Gross-Migdal}.  This limit is
characterized by the simultaneous conditions
\begin{equation}
N \to \infty, \qquad g \to g_c \, ,
\label{double-scaling}
\end{equation}
where $g$ is a typical self-coupling and $g_c$ is the location of some
large-$N$ phase transition.  The limits are however not independent.
In order to get nontrivial results, one is bound to tune the two
conditions (\ref{double-scaling}) in such a way that the combination
\begin{equation}
x = (g-g_c) N^{2/\gamma_1}
\end{equation}
is kept finite and fixed.  $\gamma_1$ is a computable critical
exponent, usually called ``string susceptibility''.  According to
Ref.\ \cite{Knizhnik-Polyakov-Zamolodchikov}, it is related to the
central charge $c$ of the model by
\begin{equation}
\gamma_1 = {1\over12}\left[25 - c + \sqrt{(1-c)(25-c)}\right].
\end{equation}

An interesting reinterpretation of the double-scaling limit relates it
to some kind of finite-size scaling in a space where $N$ plays the
r\^ole of the physical dimension $L$
\cite{Damgaard-Heller,Carlson,Brezin-ZinnJustin-RG}.  Research in this
field has exploded in many directions.  A wide review reflecting the
state of the art as of the year 1993 appeared in the already-mentioned
volume by Brezin and Wadia \cite{Brezin-Wadia}.  Here we shall only
consider those results that are relevant to our more restricted
subject.

\section{Unitary matrices}
\label{unitary-matrices}

\subsection{General features of unitary-matrix models}
\label{general-unitary-matrices}

Under the header of unitary-matrix models we class all the systems
characterized by dynamical degrees of freedom that may be expressed in
terms of the matrix representations of the unitary groups ${\rm U}(N)$
or special unitary groups ${\rm SU}(N)$ and by interactions enjoying a
global or local ${\rm U}(N)_L \times {\rm U}(N)_R$ symmetry.
Typically we shall consider lattice models, with no restriction on the
lattice structure and on the number of lattice points, ranging from 1
(single-matrix problems) to infinity (infinite-volume limit) in an
arbitrary number of dimensions.

In the field-theoretical interpretation, i.e., when considering models
in infinite volume and in proximity of a fixed point of some (properly
defined) renormalization group transformation, such models will have a
continuum counterpart, which in turn shall involve unitary-matrix
valued fields in the case of spin models, while for gauge models the
natural continuum representation will be in terms of hermitian matrix
(gauge) fields.

A common feature of all unitary-matrix models will be the
group-theoretical properties of the functional integration measure:
for each dynamical variable the natural integration procedure is based
on the left- and right-invariant Haar measure

\begin{equation}
\d\mu(U) = \d\mu(UV) = \d\mu(VU), \qquad
\int \d\mu(U) = 1.
\label{haar}
\end{equation}

An explicit use of the invariance properties of the measure and of the
interactions (gauge fixing) can sometimes lead to formulations of the
models where some of the symmetries are not apparent.  Global ${\rm
  U}(N)$ invariance is however always assumed, and the interactions,
as well as all physically interesting observables, may be expressed in
terms of invariant functions.

It is convenient to introduce some definitions and notations.  An
arbitrary matrix representation of the unitary group ${\rm U}(N)$ is
denoted by ${\cal D}_{ab}^{(r)}(U)$.  The characters and dimensions of
irreducible representations are $\chi_{(r)}(U) = {\cal
D}_{aa}^{(r)}(U)$ and $d_{(r)}$ respectively.  $(r)$ is characterized
by two set of decreasing positive integers $\{l\} = l_1,...l_s$ and
$\{m\} = m_1,...,m_t$.  We may define the ordered set of integers
$\{\lambda\} = \lambda_1,...,\lambda_N$ by the relationships
\begin{eqnarray}
\lambda_k &=& l_k,\ (k=1,...,s), \quad
\lambda_k  =  0,\ (k=s+1,...,N-t), \nonumber \\
\lambda_k &=& -m_{N-k+1},\ (k=N-t+1,...,N).
\label{lambda}
\end{eqnarray}
It is then possible to write down explicit expressions for all
characters and dimensions, once the eigenvalues $\exp\I\phi_i$ of the
matrix $U$ are known:
\begin{eqnarray}
\chi_{(\lambda)}(U) &=& {\det\Vert\exp\{\I\phi_i(\lambda_j+N-j)\}\Vert
\over \det\Vert\exp\{\I\phi_i(N-j)\}\Vert}, \\ 
d_{(\lambda)} &=& {\prod_{i<j}(\lambda_i-\lambda_j+j-i) \over 
\prod_{i<j} (j-i)} = \chi_{(\lambda)}(1).
\label{chi-d}
\end{eqnarray}

The general form of the orthogonality relations is
\begin{equation}
\int \d\mu(U) \, {\cal D}_{ab}^{(r)}(U) \, {\cal D}_{cd}^{(s)\,*}(U) =
{1\over d_{(r)}} \, \delta_{r,s}\,\delta_{a,c}\,\delta_{b,d} \,.
\label{ortho}
\end{equation}
Further relations can be found in Ref.\ \cite{Itzykson-Zuber}.

The matrix $U_{ab}$ itself coincides with the fundamental
representation $(1)$ of the group, and enjoys the properties
\begin{equation}
\chi_{(1)}(U) = \Tr U, \qquad
d_{(1)} = N, \qquad
\sum_a U_{ab}U_{ac}^* = \delta_{bc} \, .
\end{equation}
The measure $\d\mu(U)$ (which we shall also denote simply by $\d U$),
when the integrand depends only on invariant combinations, may be
expressed in terms of the eigenvalues \cite{Mehta-book}.

\subsection{Chiral models and lattice gauge theories}
\label{chiral-and-lattice}

Unitary matrix models defined on a lattice can be divided into two
major groups, according to the geometric and algebraic properties of
the dynamical variables: when the fields are defined in association
with lattice sites, and the symmetry group is global, i.e., a single 
${\rm U}(N)_L \times {\rm U}(N)_R$ transformation is applied to all
fields, we are considering a spin model (principal chiral
model); in turn, when the dynamical variables are defined on the links
of the lattice and the symmetry is local, i.e., a different
transformation for each site of the lattice may be performed, we are
dealing with a gauge model (lattice gauge theory).
As we shall see, these two classes are not unrelated to each other: an
analogy between $d$-dimensional chiral models and $2d$-dimensional
gauge theories can be found according to the following correspondence
table \cite{Green-Samuel-chiral}:
\begin{eqnarray*}
\begin{tabular}{c@{\quad}c}
{\bf spin} & {\bf gauge} \\
site, link & link, plaquette \\
loop & surface \\
length & area \\
mass & string tension \\
two-point correlation & Wilson loop \\
\end{tabular}
\end{eqnarray*}
While this correspondence in arbitrary dimensions is by no means
rigorous, there is some evidence supporting the analogy.

In the case $d=1$, which we shall carefully discuss later, one can
prove an identity between the partition function (and appropriate
correlation functions) of the two-dimensional lattice gauge theory and
the corresponding quantities of the one-dimensional principal chiral
model.  Both theories are exactly solvable, both on the lattice and in
the continuum limit, and the correspondence can be explicitly shown.

Approximate real-space renormalization recursion relations obtained by
Migdal \cite{Migdal} are identical for $d$-dimensional chiral models
and $2d$-dimensional gauge models.

The two-dimensional chiral model and the (phenomenologically
interesting) four-dimensional non-Abelian gauge theory share the
property of asymptotic freedom and dynamical generation of a mass
scale.  In both models these properties are absent in the Abelian case
(${\rm XY}$ model and ${\rm U}(1)$ gauge theory respectively), which
shows no coupling-constant renormalization in perturbation theory.

The structure of the high-temperature expansion and of the
Schwinger-Dyson equations is quite similar in the two models.

It will be especially interesting for our purposes to investigate the
Schwinger-Dyson equations of unitary-matrix models and discuss the
peculiar properties of their large-$N$ limit.

\subsection{Schwinger-Dyson equations in the large-$N$ limit} 
\label{SD-largeN}

In order to make our analysis more concrete, we must at this stage
consider specific forms of interactions among unitary matrices, both
in the spin and in the gauge models.  The most dramatic restriction
that we are going to impose on the lattice action is the condition of
considering only nearest-neighbor interactions.  The origin of this
restriction is mainly practical, because non nearest-neighbor
interactions lead to less tractable problems.  We assume that, for the
systems we are interested in, it will always be possible to find a
lattice representation in terms of nearest-neighbor interactions
within the universality class.

Let us denote by $x$ an arbitrary lattice site, and by $x,\mu$ an
arbitrary lattice link originating in the site $x$ and ending in the
site $x+\mu$: $\mu$ is one of the $d$ positive directions in a
$d$-dimensional hypercubic lattice.
A plaquette is identified by the label $x,\mu,\nu$, where the
directions $\mu$ and $\nu$ ($\mu\ne\nu$) specify the plane where the
plaquette lies.  The dynamical variables (which we label by $U$ in the
general case) are site variables $U_x$ in spin models and link
variables $U_{x,\mu}$ in gauge models.

The general expression for the partition function is
\begin{equation}
Z = \int\prod\d\mu(U) \exp[-\beta S(U)],
\end{equation}
where $\beta$ is the inverse temperature (inverse coupling) and the
integration is extended to all dynamical variables.  The action $S(U)$
must be a function enjoying the property of extensivity and of (global
and local) group invariance, and respect the symmetry of the lattice.
Adding the requisite that the interactions involve only nearest
neighbors, we find that a generic contribution to the action of spin
models must be proportional to
\begin{equation}
\sum_{x,\mu} \chi_{(r)}(U_x U^\dagger_{x+\mu}) + \hbox{h.c.} \, ,
\label{S-spin}
\end{equation}
and for gauge models to
\begin{equation}
\sum_{x,\mu,\nu} \chi_{(r)}(U_{x,\mu} U_{x+\mu,\nu}
U^\dagger_{x+\nu,\mu} U^\dagger_{x,\nu}) + \hbox{h.c.} \, ,
\label{S-gauge}
\end{equation}
where $(r)$ is in principle arbitrary, and the summation is extended
to all oriented links of the lattice in the spin case, to all the
oriented plaquettes in the gauge case.
In practice we shall mostly focus on the simplest possible choice,
corresponding to the fundamental representation.  In order to reflect 
the extensivity of the action, i.e., the proportionality to the number
of space and internal degrees of freedom, it will be convenient to
adopt the normalizations
\begin{eqnarray}
S(U) &=& -\sum_{x,\mu} N(\Tr U_x U^\dagger_{x+\mu} + \hbox{h.c.}) 
\qquad{\rm(spin)},
\label{action-spin} \\
S(U) &=& -\sum_{x,\mu,\nu} N(\Tr U_{x,\mu} U_{x+\mu,\nu}
 U^\dagger_{x+\nu,\mu} U^\dagger_{x,\nu} + \hbox{h.c.})
\qquad{\rm(gauge)}.
\label{action-gauge}
\end{eqnarray}

Once the lattice action is fixed, it is easy to obtain sets of
Schwinger-Dyson equations relating the correlation functions of the
models.  These are the quantum field equations and solving them
corresponds to finding a complete solution of a model.  It is
extremely important to notice the simplifications occurring in the
Schwinger-Dyson equations when the large-$N$ limit is considered.
These simplifications are such to allow, in selected cases, explicit
solutions to the equations.

Before proceeding to a derivation of the equations, we must
preliminarily identify the sets of correlation functions we are
interested in.  For obvious reasons, these correlations must involve
the dynamical fields at arbitrary space distances, and must be
invariant under the symmetry group of the model.  Without pretending
to achieve full generality, we may restrict our attention to such
typical objects as the invariant correlation functions of a spin model
\begin{equation}
G^{(n)}(x_1,y_1,...,x_n,y_n) = {1\over N}\left<\Tr\prod_{i=1}^n
    U_{x_i} U^\dagger_{y_i}\right>
\label{corr-spin}
\end{equation}
and to the so-called Wilson loops of a gauge model
\begin{equation}
W({\cal C}) = {1\over N}\left<\Tr\prod_{l\in{\cal C}} U_l\right>,
\label{corr-gauge}
\end{equation}
where ${\cal C}$ is a closed arbitrary walk on the lattice, and 
$\prod_{l\in{\cal C}}$ is the ordered product over all the links along
the walk.  It is worth stressing that the action itself is a sum of
elementary Green's functions (elementary Wilson loops).

More general invariant correlation functions may involve expectation
values of products of invariant operators similar to those appearing
in the r.h.s.\ of Eqs.\ (\ref{corr-spin}) and (\ref{corr-gauge}).  The
already mentioned property of factorization allows us to express the
large-$N$ limit expectation value of such products as a product of
expectation values of the individual operators.  As a consequence, the
large-$N$ form of the Schwinger-Dyson equations is a (generally
infinite) set of equations involving only the above-defined
quantities.

For sake of clarity and completeness, we present the explicit
large-$N$ form of the Schwinger-Dyson equations for the models
described by the standard actions (\ref{action-spin}) and
(\ref{action-gauge}).  For principal chiral models
\cite{GonzalezArroyo-Okawa-reduced},
\begin{eqnarray}
0 &=& G^{(n)}(x_1,y_1,...,x_n,y_n) \nonumber \\
  &+& \beta\sum_\mu\Bigl[G^{(n+1)}(x_1,x_1+\mu,x_1,y_1,...,x_n,y_n) 
      - G^{(n)}(x_1+\mu,y_1,...,x_n,y_n)\Bigr] \nonumber \\
  &+& \sum_{s=2}^n \Bigl[\delta_{x_1,x_s} \,
      G^{(s-1)}(x_1,y_1,...,x_{s-1},y_{s-1}) \,
      G^{(n-s+1)}(x_s,y_s,...,x_n,y_n) \nonumber \\
  &&\quad -\,
      \delta_{x_1,y_s} \,
      G^{(s)}(x_1,y_1,...,x_s,y_s) \,
      G^{(n-s)}(x_{s+1},y_{s+1},...,x_n,y_n)\Bigr].
\label{princ-SD}
\end{eqnarray}
For lattice gauge theories \cite{Makeenko-Migdal-exact,Wadia-study},
\begin{eqnarray}
\beta\Bigl[\sum_\mu W({\cal C}_{x,\mu\nu}) - 
                    W({\cal C}_{x-\mu,\mu\nu})\Bigr] =
\sum_{y\in{\cal C}} \delta_{x,y}\,W({\cal C}_{x,y})\,W({\cal C}_{y,x}),
\label{Migdal-Makeenko}
\end{eqnarray}
where $W({\cal C}_{x,\mu\nu})$ is obtained by replacing $U_{x,\nu}$ 
with $U_{x,\mu} U_{x+\mu,\nu} U^\dagger_{x+\nu,\mu}$ in the loop
${\cal C}$, and ${\cal C}_{x,y}$, ${\cal C}_{y,x}$ are the sub-loops
obtained by splitting ${\cal C}$ at the intersection point, including
the ``trivial'' splitting.  Eqs.\ (\ref{Migdal-Makeenko}) are commonly
known as the lattice Migdal-Makeenko equations.  The derivation of the
Schwinger-Dyson equations is obtained by performing infinitesimal
variations of the integrand in the functional integral representation
of expectation values and exploiting invariance of the measure.

\subsection{Survey of different approaches}
\label{approach-survey}

Schwinger-Dyson equations are  the starting point for most techniques
aiming at the explicit evaluation of large-$N$ vacuum expectation
values for nontrivial unitary-matrix models.  The form exhibited in
Eqs.\ (\ref{princ-SD}) and (\ref{Migdal-Makeenko}) involves in
principle an infinite set of variables, and it is therefore not
immediately useful to the purpose of finding explicit solutions.

Successful attempts to solve large-$N$ matrix systems have in general
been based on finding reformulations of Schwinger-Dyson equations
involving more restricted sets of variables and more compact
representations (collective fields).  As a matter of fact, in most
cases it turned out to be convenient to define generating functions,
whose moments are the correlations we are interested in, and whose
properties are usually related to those of the eigenvalue
distributions for properly chosen covariant combinations of matrix
fields.

By ``covariant combination'' we mean a matrix-valued variable whose
eigenvalues are left invariant under a general ${\rm SU}(N) \times
{\rm SU}(N)$ transformation of the Lagrangian fields.  Such objects
are typically those appearing in the r.h.s.\ of Eqs.\ (\ref{corr-spin})
and (\ref{corr-gauge}) {\em before} the trace operation is performed.
Under the ${\rm SU}(N) \times {\rm SU}(N)$ transformation $U \to
VUW^\dagger$, these operators transform accordingly to ${\cal O} \to
V{\cal O}V^\dagger$, and therefore their eigenvalue spectrum is left
unchanged.

Without belaboring on the details (some of which will however be
exhibited in the discussion of the single-link integral presented in
Sect.\ \ref{single-link}), we only want to mention that the approach
based on extracting appropriate Schwinger-Dyson equations for the
generating functions is essentially algebraic in nature, involving
weighted sums of infinite sets of equations in the form
(\ref{princ-SD}) or (\ref{Migdal-Makeenko}), identification of the
relevant functions, and resolution of the resulting algebraic
equations, where usually a number of free parameters appear, whose
values are fixed by boundary and/or asymptotic conditions and
analyticity constraints.  
The approach based on direct replacement of the eigenvalue
distributions in the functional integral and the minimization of the
resulting effective action leads in turn to integral equations which
may be solved by more or less straightforward techniques.
These two approaches are however intimately related, since the
eigenvalue density is usually connected with the discontinuity along
some cut in the complex-plane extension of the generating function,
and one may easily establish a step-by-step correspondence between the
algebraic and functional approach.

Let us finally mention that the procedure based on introducing
invariant degrees of freedom and eigenvalue density operators has been
formalized by Jevicki and Sakita
\cite{Jevicki-Sakita-collective,Jevicki-Sakita-euclidean} in terms of
a ``quantum collective field theory'', whose equations of motion are
the Schwinger-Dyson equations relevant to the problem at hand.

A quite different application of the Schwinger-Dyson equations is
based on the strong-coupling properties of the correlation functions.
In the strong-coupling domain, expectation values are usually analytic
in the coupling $\beta$ within some positive convergence radius, and
their boundary value at $\beta=0$ can easily be evaluated.  As a
consequence, it is formally possible to solve Eqs.\ (\ref{princ-SD})
and (\ref{Migdal-Makeenko}) in terms of strong-coupling series by
sheer iteration of the equations.  This procedure may in practice turn
out to be too cumbersome for practical purposes; however, in some
circumstances, it may lead to rather good approximations
\cite{Marchesini-loop,Marchesini-Onofri-convergence} and even to a
complete strong-coupling solution.  Continuation to the weak-coupling
domain is however a rather nontrivial task.

As a special application of the strong-coupling approach, we must
mention the attempt (pioneered by Kazakov, Kozhamkulov and Migdal
\cite{Kazakov-Kozhamkulov-Migdal}) to construct an effective action
for the invariant degrees of freedom by means of a modified
strong-coupling expansion, and explore the weak-coupling regime by
solving the saddle-point equations of the resulting action.  This
technique might be successful at least in predicting the location and
features of the large-$N$ phase transition which is relevant to many
physical problems, as mentioned in Sect.\ \ref{introduction}.

A numerical approach to large-$N$ lattice Schwinger-Dyson equations
based on the minimization of an effective large-$N$ Fokker-Plank
potential and suited for the weak-coupling regime was proposed by
Rodrigues \cite{Rodrigues-numerical}.

Another relevant application of the Schwinger-Dyson equations is found
in the realm of the so-called ``reduced'' models.  These models, whose
prototype is the Eguchi-Kawai formulation of strong-coupling large-$N$
lattice gauge theories \cite{Eguchi-Kawai-reduction}, are based on the
physical intuition that, in the absence of fluctuations, due to
translation invariance, the space extension of the lattice must be
essentially irrelevant in the large-$N$ limit, since all invariant
physics must be already contained in the expectation values of
(properly chosen) purely local variables.  More precisely, one might
say that, when $N\to\infty$, the ${\rm SU}(N)$ group becomes so large
that it accommodates the full Poincar\`e group as a subgroup, and in
particular it should be possible to find representations of the
translation and rotation operators among the elements of ${\rm
SU}(N)$.  As a consequence, one must be able to reformulate the full
theory in terms of a finite number of matrix field variables defined
at a single space-time site (or on the $d$ links emerging from the
site in the case of a lattice gauge theory) and of the above-mentioned
representations of the translation group.  This reformulation is
called ``twisted Eguchi-Kawai'' reduced version of the theory
\cite{Eguchi-Nakayama-simplification,GonzalezArroyo-Okawa-EK}.

We shall spend a few more words on the reduced models in Sect.\ 
\ref{principal-chiral}.  Moreover, a very good review of their
properties has already appeared many years ago \cite{Das-review}.  In
this context, we must only mention that the actual check of validity
of the reduction procedure is based on deriving the Schwinger-Dyson
equations of the reduced model and comparing them with the
Schwinger-Dyson equations of the original model.  Usually the
equivalence is apparent already at a superficial level when
na{\"\i}vely applying to correlation functions of the reduced model
the symmetry properties of the action itself.  This procedure however
requires some attention, since the limit of infinitely many degrees of
freedom within the group itself allows the possibility of spontaneous
breakdown of some of the symmetries which would be preserved for any
finite value of $N$.  In this context, we recall once more that large
$N$ is a thermodynamical limit: $N$ must go to infinity before any
other limit is considered, and sometimes the limiting procedures do
not commute.  It is trivial to recognize that, when the
strong-coupling phase is considered, symmetries are unbroken, and the
equivalence between original and reduced model may be established
without further ado.  Problems may occur in the weak-coupling side of
a large-$N$ phase transition.

An unrelated and essentially numeric approach to solving the large-$N$
limit of lattice matrix models is the coherent state variational
algorithm introduced by Yaffe and coworkers
\cite{Brown-Yaffe,Dickens-Lindqwister-Somsky-Yaffe}.  We refer to the
original papers for a presentation of the results that may be obtained
by this approach.

\section{The single-link integral}
\label{single-link}

\subsection{The single-link integral in external field: finite-$N$
solution}
\label{single-link-finite-N}

All exact and approximate methods of evaluation of the functional
integrals related to unitary-matrix models must in principle face the
problem of performing the simplest of all relevant integrations: the
single-link integral.  The utmost importance of such an evaluation
makes it proper to devote to it an extended discussion, which will
also give us the opportunity of discussing in a prototype example the
different techniques that may be applied to the models we are
interested in.

A quite general class of single-link integrals may be introduced by
defining 
\begin{equation}
Z(A^\dagger A) = \int\d U
\exp\bigl[N\Tr(A^\dagger U + U^\dagger A)\bigr],
\label{one-link}
\end{equation}
where as usual $U$ is an element of the group ${\rm U}(N)$ and $A$ is
now an arbitrary $N\times N$ matrix.  The ${\rm U}(N)$ invariance of
the Haar measure implies that the one link integral (\ref{one-link})
must depend only on the eigenvalues of the Hermitian matrix $A^\dagger
A$, which we shall denote by $x_1,...,x_N$.  The function
$Z(x_1,...,x_N)$ must satisfy a Schwinger-Dyson equation: restricting
the variables to the ${\rm U}(N)$ singlet subspace, the
Schwinger-Dyson equation was shown to be equivalent to the partial
differential equation \cite{Brower-Nauenberg,Brezin-Gross}
\begin{eqnarray}
{1\over N^2}\, x_k\,{\partial^2 Z\over\partial x_k^2} + 
{1\over N}\,{\partial Z\over\partial x_k} + 
{1\over N^2} \sum_{s \ne k} {x_s\over x_k-x_s}
\left({\partial Z\over\partial x_k} - 
      {\partial Z\over\partial x_s}\right)
= Z, \nonumber \\
\label{SD-Z}
\end{eqnarray}
with the boundary condition $Z(0,...,0) = 1$ and the request that $Z$
be completely symmetric under exchange of the $x_i$.

It is convenient to reformulate the equation in terms of the new
variables $z_k = 2N\sqrt{x_k}$, and to parameterize the solution in
terms of the completely antisymmetric function 
${\hat Z}(z_1,...,z_N)$ by defining
\begin{equation}
Z(z) = {{\hat Z}(z)\over\prod_{i<j}(z_i^2-z_j^2)}\,.
\end{equation}
The equation satisfied by $\hat Z$ can be shown to reduce to
\begin{eqnarray}
&&\Biggl[ \sum_k z_k^2\,{\partial^2\over\partial z_k^2} +
    (3-2N) \sum_k z_k\,{\partial\over\partial z_k} - \sum_k z_k^2 + 
    {2\over3}\,N(N-1)(N-2)\Biggr] {\hat Z} = 0. \nonumber \\
\label{Zhat}
\end{eqnarray}
Eq.\ (\ref{Zhat}) has the structure of a fermionic many-body
Schr\"odinger equation.  With some ingenuity it may be solved in the
form of a Slater determinant of fermion wavefunctions.  In conclusion,
we obtain, after proper renormalization
\cite{Brower-Rossi-Tan-chains} (see also \cite{Gaudin-Mello}),
\begin{equation}
Z(z_1,...,z_N) = 2^{N(N-1)/2} \Biggl(\prod_{k=0}^{N-1} k!\Biggr)
   {\det\Vert z_j^{i-1} I_{i-1}(z_j)\Vert \over
    \det\Vert z_j^{2(i-1)}\Vert},
\label{Z}
\end{equation}
where $I_i(z)$ is the modified Bessel function.  Eq.\ (\ref{Z}) is
therefore a representation of the single-link integral in external
field for arbitrary ${\rm U}(N)$ groups.  By taking proper derivatives
with respect to its arguments one may in principle reconstruct all the
cumulants for the group integration of an arbitrary string of
(uncontracted) matrices \cite{Samuel-integrals,Bars}.

Some special limits of the general expression (\ref{Z}) may prove
useful.  Let us first of all consider the case when $A$ is
proportional to the identity matrix: $A = a 1$ and therefore
$z_i = 2Na$ and
\begin{equation}
Z(2Na,...,2Na) = \det\Vert I_{i-j}(2Na)\Vert.
\end{equation}
As we shall see, this is exactly Bars' and Green's solution for ${\rm
U}(N)$ lattice gauge theory in two dimensions
\cite{Bars-Green-largeN}.

When only one eigenvalue of $A$ is different from zero the result is
\begin{equation}
Z(2Na,0,...,0) = (N-1)!\,(Na)^{1-N}\,I_{N-1}(2Na).
\end{equation}
The large-$N$ limit will be discussed in the next subsection.

\subsection{The external field problem: large-$N$ limit}
\label{external-field-large-N}

For our purposes it is extremely important to extract the limiting
form of Eq.\ (\ref{Z}) when $N\to\infty$.  In principle, it is a very
involved problem, since the dependence on $N$ comes not only through
the $z_i$ but also from the dimension of the matrices whose
determinant we must evaluate.  It is however possible to obtain the
limit, either by solving separately the large-$N$ version of
Eq.\ (\ref{SD-Z}), or by directly manipulating Eq.\ (\ref{Z}).

In the first approach, we introduce the large-$N$ parameterization
\begin{equation}
Z = \exp NW,
\end{equation}
where $W$ is now proportional to $N$; we then obtain from
Eq.\ (\ref{SD-Z}), dropping second-derivative terms that are manifestly
depressed in the large-$N$ limit \cite{Brezin-Gross},
\begin{equation}
x_k\left(\partial W\over\partial x_k\right)^{\!\!2} + 
{\partial W\over\partial x_k} +
{1\over N}\sum_{s\ne k}{x_s\over x_s-x_k}
\left({\partial W\over\partial x_s} - 
    {\partial W\over\partial x_k}\right) = 1.
\label{SD-Z-largeN}
\end{equation}
It is possible to show that in the large-$N$ limit
Eq.\ (\ref{SD-Z-largeN}) admits solutions, which can be parameterized
by the expression
\begin{equation}
{\partial W\over\partial x_k} = {1\over\sqrt{x_k+c}}
\left[1 - {1\over2N}\sum_s{1\over \sqrt{x_k+c} + \sqrt{x_s+c}}\right],
\qquad c \ge 0.
\label{SD-Z-largeN-sol}
\end{equation}
Substitution of Eq.\ (\ref{SD-Z-largeN-sol}) into
Eq.\ (\ref{SD-Z-largeN}) and some algebraic manipulation lead to the
consistency condition
\begin{equation}
c\left[{1\over2N}\sum_s{1\over\sqrt{x_s+c}} - 1\right] = 0,
\end{equation}
which in turn admits two possible solutions:

{\it a}) $c$ is determined by the condition
\begin{equation}
{1\over2N}\sum_s{1\over\sqrt{x_s+c}} = 1,
\label{SD-Z-strong-cond}
\end{equation}
implying $c\le{1\over4}$; this is a ``strong coupling'' phase,
requiring that the eigenvalues satisfy the bound
\begin{equation}
{1\over2N}\sum_s{1\over \sqrt{x_s}} \ge 1,
\label{SD-Z-strong-reg}
\end{equation}
i.e., at least some of the $x_s$ are sufficiently small;

{\it b}) when 
\begin{equation}
{1\over2N}\sum_s{1\over\sqrt{x_s}} \le 1,
\label{SD-Z-weak-cond}
\end{equation}
then the solution corresponds to the choice $c=0$; this is a ``weak
coupling'' phase, and all eigenvalues are large enough.

Direct integration of Eq.\ (\ref{SD-Z-largeN-sol}) with proper boundary
conditions leads to the large-$N$ result \cite{Brezin-Gross}
\begin{eqnarray}
W(x) &=& 2\sum_k\sqrt{x_k+c} - {1\over2N}\sum_{k,s}\log
\bigl(\sqrt{x_k+c} + \sqrt{x_s+c}\bigr) 
- Nc - {3\over4}N, \nonumber \\
\label{SD-W}
\end{eqnarray}
which must be supplemented with Eq.\ (\ref{SD-Z-strong-cond}) in the
strong-coupling regime (\ref{SD-Z-strong-reg}), while $c=0$ reproduces
the weak-coupling result by Brower and Nauenberg.  Amazingly enough,
setting $c=0$ in Eq.\ (\ref{SD-W}) one obtains the n{a\"\i}ve
one-loop estimate of the functional integral, which turns out to be
exact in this specific instance.

It is possible to check that Eq.\ (\ref{SD-W}) is reproduced
by carefully taking the large-$N$ limit of Eq.\ (\ref{Z}), which
requires use of the following asymptotic limits of Bessel functions
\cite{Brower-Rossi-Tan-chains}
\begin{eqnarray}
k! \left(2\over z\right)^{\!\!k} I_k(z) &&\goto_{z\to\infty}
\left[{1\over2}\left(1+\sqrt{1+{z^2\over k^2}}\right)\right]^{1-k}
\left(1+{z^2\over k^2}\right)^{\!\!-1/4} \nonumber \\ 
&&\quad\times\,\exp\left(\sqrt{k^2+z^2} - k\right) \qquad 
\hbox{(strong coupling)}, \\ 
I_k(z)\, &&\approx {1\over\sqrt{2\pi z}} \exp z \qquad
\hbox{(weak coupling)}.
\end{eqnarray}

An essential feature of Eq.\ (\ref{SD-W}) is the appearance of two
different phases in the large-$N$ limit of the single-link integral.
Such a transition would be mathematically impossible for any finite
value of $N$; however it affects the large-$N$ behavior of all
unitary-matrix models and gives rise to a number of interesting
phenomena.  A straightforward analysis of Eq.\ (\ref{SD-W}) shows that
the transition point corresponds to the condition
\begin{equation}
t \equiv {1\over2N}\sum_s{1\over\sqrt{x_s}} = 1.
\end{equation}

It is also possible to evaluate the difference between the strong- and
weak-coupling phases of $W$ in the neighborhood of $t=1$, finding the
relationship \cite{Brezin-Gross}
\begin{equation}
W_{\rm strong} - W_{\rm weak} \sim (t-1)^3.
\end{equation}
As a consequence, we may classify this phenomenon as a ``third order
phase transition''.

\subsection{The properties of the determinant}
\label{properties-determinant}

The large-$N$ factorization of invariant amplitudes is a
well-estab\-lished property of products of operators defined starting
from the fundamental representation of the symmetry group.  Operators
corresponding to highly nontrivial representations may show a more
involved pattern of behavior in the large-$N$ limit.  Especially
relevant from this point of view are the properties of determinants of
covariant combinations of fields
\cite{Green-Samuel-chiral,Green-Samuel-un2}; we will consider the
quantities
\begin{equation}
\Delta(x) = \det\left[U_0 U^\dagger_x\right]
\end{equation}
for lattice chiral models and
\begin{equation}
\Delta({\cal C}) = \det\prod_{l\in{\cal C}} U_l
\end{equation}
for lattice gauge theories.

The expectation values of these operators may act as an order
parameter for the large-$N$ phase transition characterizing the class
of models we are taking into consideration.  Indeed the determinant
picks up the phase characterizing the ${\rm U}(1)$ subgroup that
constitutes the center of ${\rm U}(N)$.  Moreover, since
\[
{\rm U}(N) \approx {\rm U}(1) \times {{\rm SU}(N)\over Z_N}\,,
\]
${\rm SU}(N) \to {\rm U}(N)$ as $N\to\infty$ because $Z_N\to{\rm
U}(1)$; therefore the determinant of the ${\rm U}(N)$ theory in the
large-$N$ limit reflects properties of the center of ${\rm SU}(N)$.

In lattice models this Abelian ${\rm U}(1)$ subgroup is not decoupled,
as it happens in the continuum theory, and therefore
$\left<\Delta\right>$ does not in general have on the lattice the
free-theory behavior it has in the continuum.

The basic properties of the determinant may be explored by focusing
once more on the external field problem we discussed above.  Let us
introduce a class of determinant operators, and define their
expectation values as \cite{Aneva-Brihaye-Rossi-pseudoAbelian}
\begin{equation}
\Delta^{(l)} = \left<\det U^l\right> =
{\int\d U \det U^l \exp[N\Tr(U^\dagger A + A^\dagger U)] \over
 \int\d U \exp[N\Tr(U^\dagger A + A^\dagger U)]} \,.
\end{equation}
In order to parameterize the ${\rm SU}(N)$ external-source integral,
besides the eigenvalues $x_i$ of $AA^\dagger$, a new external
parameter must be introduced, that couples to the determinant:
\begin{equation}
\theta = {\I\over2N}(\log\det A^\dagger - \log\det A).
\label{theta-def}
\end{equation}
Because of the symmetry properties, $\Delta^{(l)}$ may only depend on
the eigenvalues $z$ and on $\theta$.
It was found that, when $U$ enjoys ${\rm U}(N)$ symmetry (with finite
$N$), 
\begin{equation}
\Delta^{(l)} = \exp(\I Nl\theta) {\hat Z_l\over\hat Z_0} \,,
\end{equation}
where $\hat Z_l$ is the solution of the following Schwinger-Dyson
equation, generalizing Eq.\ (\ref{Zhat}):
\begin{eqnarray}
&&{1\over N}\Biggl[\sum_k z^2_k\,{\partial^2\over\partial z_k^2} +
    (3-2N)\sum_k z_k\,{\partial\over\partial z_k} - \sum_k z^2_k 
\nonumber \\ && \qquad+\,
    {2\over3}N(N-1)(N-2)\Biggr]\hat Z_l = l^2 \hat Z_l;
\label{SD-hatZl}
\end{eqnarray}
$\hat Z_l$ satisfy the property
\begin{equation}
\hat Z_l = \Biggl(\prod_k z_k\Biggr)^{\!\!|l|}
\Biggl(\prod_k {1\over z_k}\,
{\partial\over\partial z_k}\Biggr)^{\!\!|l|}
\hat Z_0 = \det\Vert z_i^{j-1} I_{j-1-l}(z_i)\Vert.
\label{Zhatl-def}
\end{equation}

When the weak-coupling condition $t\equiv\sum_k 1/z_k\le1$ is
satisfied, the leading contribution to the large-$N$ limit of all
$\hat Z_l$ is the same:
\begin{equation}
\hat Z_l \to \hat Z^{(\infty)} =
\exp\left[\sum_k z_k - {1\over2}\sum_k\log 2\pi z_k +
    \sum_{i<k}\log(z_i-z_k)\right].
\label{hatZl}
\end{equation}
In order to determine the large-$N$ limit of $\Delta^{(l)}$, one
therefore needs to compute the $O(1)$ factor in front of the
exponentially growing term (\ref{hatZl}).  It is convenient to define
\begin{equation}
X_l = {\hat Z_l\over\hat Z^{(\infty)}} \,,
\label{Xl-def}
\end{equation}
whose Schwinger-Dyson equation may be extracted from
Eq.\ (\ref{SD-hatZl}) and takes the form
\begin{eqnarray}
&&{1\over N}\left[\sum_k z^2_k\,{\partial^2 X_l\over\partial z_k^2} +
    2 \sum_k z^2_k\,{\partial X_l\over\partial z_k} +
    \sum_{k\ne i} {z_k z_i\over z_k-z_i}
        \left({\partial X_l\over\partial z_k} - 
              {\partial X_l\over\partial z_i}\right)\right] 
\nonumber \\ =\,&&
\left(l^2-{1\over4}\right) X_l \,.
\label{SD-Xl}
\end{eqnarray}

Let us introduce the large-$N$ Ansatz
\begin{equation}
X_l = X_l(t),
\end{equation}
reducing Eq.\ (\ref{SD-Xl}) to
\begin{equation}
{1\over N} \sum_k{1\over z^2_k}\,{\d^2X_l\over\d t^2} +
2(t-1){\d X_l\over\d t} = \left(l^2-{1\over4}\right) X_l \,.
\label{SD-Xl-reduced}
\end{equation}
Removing terms that are depressed by two powers of $1/N$, we are left
with a consistent equation whose solution is
\begin{equation}
X_l = (1-t)^{{1\over2}(l^2-{1\over4})}.
\label{SD-Xl-sol}
\end{equation}
Finally we can compute the weak-coupling large-$N$ limit of
$\Delta^{(l)}$:
\begin{equation}
\Delta^{(l)} \goto_{N\to\infty} \exp(\I Nl\theta) \,
    (1-t)^{{1\over2}l^2}, \qquad t \le 1.
\end{equation}

From the standard strong-coupling expansion we may show that
\begin{equation}
\Delta^{(l)} \goto_{N\to\infty} 0
\qquad \hbox{when} \quad t \ge 1.
\end{equation}

An explicit evaluation, starting from the exact expression
(\ref{Zhatl-def}), expanded in powers of $1/z_k$ for arbitrary $N$,
allows us to show that the quantities $\hat Z_l$ may be obtained from
Eqs.\ (\ref{hatZl}) and (\ref{Xl-def}) by expanding
Eq.\ (\ref{SD-Xl-sol}) up to 2nd order in $t$ with no $O(1/N^2)$
corrections.  $\Delta^{(l)}$ according to this result violate
factorization; in turn, they take the value which would be predicted
by an effective Gaussian theory governing the ${\rm U}(1)$ phase of
the field $U$.

\subsection{Applications to mean field and strong coupling}
\label{mean-field+strong-coupling}

The single-link external-field integral has a natural domain of
application in two important methods of investigation of lattice field
theories: mean-field and strong-coupling expansion.  Extended papers
and review articles have been devoted in the past to these topics
(cfr.\ Ref.\ \cite{Drouffe-Zuber} and references therein), and we
shall therefore focus only on those results that are specific to the
large-$N$ limit and to the $1/N$ expansion.

Let us first address the issue of the mean-field analysis, considering
for sake of definiteness the case of $d$-dimensional chiral models,
but keeping in mind that most results can be generalized in an
essentially straightforward manner to lattice gauge theories.  The
starting point of the mean-field technique is the application of the
random field transform to the functional integral:
\begin{eqnarray}
Z_N &=& \int\d U_n \exp\Biggl\{N\beta\sum_{n,\mu}
        \Tr\left(U_n U^\dagger_{n+\mu}
            + U_{n+\mu} U^\dagger_n\right)\Biggr\} 
\nonumber \\
&=& \int\d V_n \d A_n \exp\Biggl\{N\beta\sum_{n,\mu}
        \Tr\left(V_n V^\dagger_{n+\mu} 
            + V_{n+\mu} V^\dagger_n\right)
\nonumber \\
&&\qquad\quad-\, N\sum_n \Tr\left(A_n V_n^\dagger
            + V_n A^\dagger_n\right) \Biggr\}
\nonumber \\
&&\qquad\times\,
    \int\d U_n \exp\left\{N \sum_{n,\mu}
        \Tr\left(A_n U_n^\dagger+ U_n A^\dagger_n\right)\right\},
\end{eqnarray}
where $V_n$ and $A_n$ are arbitrary complex $N{\times}N$ matrices.
Therefore the integration over $U_n$ is just the single-link integral
we discussed above.  As a consequence, the original chiral model is
formally equivalent to a theory of complex matrices with effective
action
\begin{eqnarray}
-{1\over N} \, S_{\rm eff}(A,V) &=& \beta\sum_{n,\mu} 
        \Tr\left(V_n V_{n+\mu}^\dagger+ V_{n+\mu} V^\dagger_n\right)
- \sum_n \Tr\left(A_n V_n^\dagger+ V_n A^\dagger_n\right)
\nonumber \\ &+&
\sum_n W(A_n A^\dagger_n).
\label{chiral-Seff}
\end{eqnarray}
The leading order in the mean-field approximation is obtained by
applying saddle-point techniques to the effective action, assuming
saddle-point values of the fields $A_n$ and $V_n$ that are
translation-invariant and proportional to the identity.

We mention that, in the case at hand, the large-$N$ saddle-point
equations in the weak-coupling phase are:
\begin{equation}
A_n = a = 2 \beta d v, \qquad V_n = v = 1 - {1 \over 4a},
\end{equation}
and they are solved by the saddle-point values
\begin{equation}
\overline a = \beta d \left(1 + \sqrt{1 - {1 \over 2 \beta d}}\right),
\qquad \overline v = {1\over2} + 
    {1\over2} \sqrt{1 - {1 \over 2 \beta d}},
\end{equation}
leading to a value of the free and internal energy
\begin{equation}
{Fd \over N^2 L} = 
\overline a - {1\over2}\log 2\overline a - {1\over2}, \qquad
{1\over2d}\,{\partial\over\partial\beta}\,{Fd \over N^2 L} = 
\overline v^2.
\end{equation}
The strong-coupling solution is trivial: $v = a = 0$, and there is a
first-order transition point at
\begin{equation}
\beta_c d = \casefr{1}{2}, \qquad \overline v_c = \casefr{1}{2},
\qquad \overline a_c = \casefr{1}{2}.
\end{equation}

One may also compute the quadratic fluctuations around the mean-field
saddle point by performing a Gaussian integral, whose quadratic form
is related to the matrix of the second derivatives of $W$ with respect
to the fields, and generate a systematic loop expansion in the
effective action (\ref{chiral-Seff}), which in turns appears to be
ordered in powers of $1/d$.  Therefore mean-field methods are
especially appropriate for the discussion of models in large space
dimensions, and not very powerful in the analysis of $d=2$ models.
The very nature of the transition cannot be taken for granted,
especially at large $N$.  However, when $d\ge3$ there is independent
evidence of a first-order phase transition for $N\ge3$.  We mention
that a detailed mean-field study of ${\rm SU}(N)$ chiral models in $d$
dimensions appeared in Refs.\ 
\cite{Kogut-Snow-Stone-mean,Brihaye-Rossi-weak}.

When willing to extend the mean-field approach, it is in general
necessary to find a systematic expansion of the functional
$W(AA^\dagger)$ in the powers of the fluctuations around the
saddle-point configurations.  Moreover, one may choose to consider not
only the large-$N$ value of the functional, but also its expansion in
powers if $1/N^2$, in order to make predictions for large but finite
values of $N$.  The expansion of $W_0$ up to fourth order in the
fluctuations was performed in Ref.\ \cite{Brihaye-Taormina-mean},
where explicit analytic results can be found.  A technique for the
weak-coupling $1/N^2$ expansion of $W$ can be found in Ref.\ 
\cite{Brihaye-Rossi-integrals}.  We quote the complete $O(1/N^4)$
result:
\begin{eqnarray}
{W \over N} = {1 \over N^2} &&\Biggl[\sum_a z_a 
    - {1\over2} \sum_{a,b}\log{z_a+z_b \over 2N} - {3\over4}\,N^2
    + \log(1-t)^{-1/8} 
\nonumber \\ &&\quad+\,
    {3\over2^7}(1-t)^{-3} \sum_a {1 \over z_a^3}\Biggr]
    + O\left(1 \over N^6\right),
\label{W-1/N2}
\end{eqnarray}
where $t = \sum_a 1/z_a$.  Eq.\ (\ref{W-1/N2}) can also be expanded in
the fluctuations around a saddle-point configuration.  Extension to
${\rm SU}(N)$ with large $N$ was also considered.  A discussion of
large-$N$ mean field for lattice gauge theories can be found in
Refs.\ \cite{Brihaye-Rossi-weak,Muller-Ruhl-mean-1,Muller-Ruhl-mean-2,%
Hasegawa-Yang-mean-1,Hasegawa-Yang-mean-2}. 

Let us now turn to a discussion of the main features of the large-$N$
strong-coupling expansion.  A preliminary consideration concerns the
fact that it is most convenient to reformulate the strong-coupling
expansion (i.e., the expansion in powers of $\beta$) into a character
expansion, which is ordered in the number of lattice steps involved in
the effective path that can be associated with each nontrivial
contribution to the functional integral.  The large-$N$ character
expansion will be discussed in greater detail in Subs.\ 
\ref{character-expansion}.  Here we only want to discuss those
features that are common to any attempt aimed at evaluating
strong-coupling series for expectation values of invariant operators
in the context of ${\rm U}(N)$ and ${\rm SU}(N)$ matrix models, with
special focus on the large-$N$ behavior of such series.

The basic ingredient of strong-coupling computations is the knowledge
of the cumulants, i.e., the connected contributions obtained
performing the invariant group integration of a string of uncontracted
$U$ and $U^\dagger$ matrices.  ${\rm U}(N)$ group invariance insures
us that these group integrals can be non-zero only if the same number
of $U$ and $U^\dagger$ matrices appear in the integrand.  ${\rm
SU}(N)$ is slightly different in this respect, and its peculiarities
will be discussed later and are not relevant to the present analysis.

It was observed a long time ago that the cumulants, whose group
structure is that of invariant tensors with the proper number of
indices, involve $N$-dependent numerical coefficients. The asymptotic
behavior of these coefficients in the large-$N$ limit was studied
first by Weingarten \cite{Weingarten-asymptotic}.  However, for finite
$N$, the coefficients written as function of $N$ are formally plagued
by the so-called DeWit-'t Hooft poles \cite{DeWit-tHooft}, that are
singularities occurring for integer values of $N$.  The highest
singular value of $N$ grows with the number $n$ of $U$ matrices
involved in the integration, and therefore for sufficiently high
orders of the series it will reach any given finite value.  A complete
description of the pole structure was presented in Ref.\ 
\cite{Samuel-integrals}; not only single poles, but also arbitrary
high-order poles appear for large enough $n$, and analyticity is
restricted to $N \ge n$.  Obviously, since group integrals are well
defined for all $n$ and $N$, this is only a pathology of the $1/N$
expansion.  Finite-$N$ results are finite, but they cannot be obtained
as a continuation of a large-$N$ strong-coupling expansion.  However,
it is possible to show that the strict $N\to\infty$ limit of the
series exists, and moreover, for sufficiently small $\beta$ and
sufficiently large $N$, the limiting series is a reasonable
approximation to the true result, all nonanalytic effects being
$O(\beta^{2N})$ in ${\rm U}(N)$ models and $O(\beta^N)$ in ${\rm
SU}(N)$ models.  As a consequence, computing the large-$N$ limit of
the strong-coupling series is meaningful and useful in order to
achieve a picture of the large-$N$ strong-coupling behavior of matrix
models, but the evaluation of $O(1/N^2)$ or higher-order corrections
in the strong-coupling phase is essentially pointless.

The large-$N$ limit of the external-field single-link integral has
been considered in detail from the point of view of the
strong-coupling expansion.  In particular, one may obtain expressions
for the coefficients of the expansion of $W$ in powers of the moments
of $AA^\dagger$: setting
\begin{equation}
\rho_n = {1 \over N}\Tr(AA^\dagger)^n, \qquad
W = \sum_{n=1}^\infty 
\sum_{\stackrel{\scriptstyle \alpha_1,...,\alpha_n}%
{\sum_k k \alpha_k = n}}
W_{\alpha_1,...,\alpha_n} 
\rho_1^{\alpha_1} ... \rho_n^{\alpha_n} \, ,
\end{equation}
one gets
\begin{equation}
W_{\alpha_1,...,\alpha_n} = 
(-1)^n {(2 n + \sum_k \alpha_k - 3)! \over (2 n)!}
\prod_k \left[-{(2 k)! \over (k!)^2}\right]^{\alpha_k}
{1 \over \alpha_k!} .
\end{equation}
Further properties of this expansion can be found in the original
reference \cite{OBrien-Zuber-note}.

A character-expansion representation of the single-link integral was
also produced for arbitrary ${\rm U}(N)$ integrals in Ref.\ 
\cite{Bars}.  Strong-coupling expansions for large-$N$ lattice gauge
theories have been analyzed in detail by Kazakov
\cite{Kazakov-pl,Kazakov-jetp}, O'Brien and Zuber
\cite{OBrien-Zuber-expansion}, and Kostov \cite{Kostov-sc}, who
proposed reinterpretations in terms of special string theories.

\subsection{The single-link integral in the adjoint representation}
\label{single-link-adjoint}

The integral introduced at the beginning of Sect.\ \ref{single-link} is
by no means the most general single-link integral one can meet in
unitary-matrix models.  As mentioned in Sect.\ \ref{unitary-matrices},
any invariant function of the $U$'s is in principle a candidate for a
lattice action.  In practice, the only case that has been considered
till now that cannot be reduced to Eq.\ (\ref{one-link}) is the
integral introduced by Itzykson and Zuber \cite{Itzykson-Zuber}
\begin{equation}
I(M_1,M_2) = \int\d U \exp\Tr(M_1 U M_2 U^\dagger),
\label{I-def}
\end{equation}
where $M_1$ and $M_2$ are arbitrary Hermitian matrices.  This is a
special instance of the single-link integral for the coupling of the
adjoint representation of $U$ to an external field.

The result, because of ${\rm U}(N)$ invariance, can only depend on the
eigenvalues $m_{1i}$ and $m_{2i}$ of the Hermitian matrices.  Several
authors \cite{Itzykson-Zuber,Mehta-integration,HarishChandra} have
independently shown that
\begin{equation}
I(M_1,M_2) = \Biggl(\prod_{p=1}^{N-1} p!\Biggr)
{\det\left\Vert\exp(m_{1i} m_{2j})\right\Vert \over
 \Delta(m_{11},...,m_{1N})\,\Delta(m_{21},...,m_{2N})},
\label{I-value-det}
\end{equation}
where $\Delta(m_1,...,m_N) = \prod_{i>j}(m_i-m_j)$ is the Vandemonde
determinant.  A series expansion for $I(M_1,M_2)$ in terms of the
characters of the unitary group takes the form
\begin{equation}
I(M_1,M_2) = \sum_{(r)} {1\over|n|!}\,
{\sigma_{(r)}\over d_{(r)}}\,\chi_{(r)}(M_1)\,\chi_{(r)}(M_2),
\end{equation}
where $\sigma_{(r)}$ is the dimension of the representation $(r)$ of
the permutation group; we will present an explicit evaluation of
$\sigma_{(r)}$ in Eq.\ (\ref{sigma-def}).  Eq.\ (\ref{I-value-det})
plays a fundamental r\^ole in the decoupling of the ``angular''
degrees of freedom when models involving complex Hermitian matrices
are considered.

An interesting development based on the use of
Eq.\ (\ref{I-value-det}) is the so-called ``induced QCD'' program,
aimed at recovering continuum large-$N$ QCD by taking proper limits in
the parameter space of the lattice Kazakov-Migdal
model \cite{Kazakov-Migdal-induced} 
\begin{equation}
S = N\sum_x \Tr V(\Phi_x) - N \sum_{x,\mu} 
\Tr(\Phi_x U_{x,\mu} \Phi_{x+\mu} U^\dagger_{x,\mu}),
\end{equation}
where $U_{x,\mu}$ is the non-Abelian gauge field and $\Phi_x$ is a
Hermitian $N \times N$ (matrix-valued) Lorentz-scalar field.  The
Itzykson-Zuber integration (\ref{I-def}) allows the elimination of the
gauge degrees of freedom and reduces the problem to studying the
interactions of Hermitian matrix fields (with self-interactions
governed by the potential $V$).  Discussion of the various related
developments is beyond the scope of the present report.  It will be
enough to say that, while one may come to the conclusion that this
model does {\em not\/} induce QCD, it is certainly related to some
very interesting (and sometimes solvable) matrix models (cfr.\ Ref.\ 
\cite{Weiss} for a review).

\section{Two-dimensional lattice Yang-Mills theory}
\label{2d-YM}

\subsection{Two-dimensional Yang-Mills theory as a single-link
integral}
\label{2d-YM-single-link}

The results presented in the previous section allow us to analyze
the simplest physical system described by a unitary-matrix model.  As
we shall see, one of the avatars of this system is a Yang-Mills theory
in two dimensions (YM$_2$), in the lattice Wilson formulation.
Notwithstanding the enormous simplifications occurring in this model
with respect to full QCD, still some nontrivial features are retained,
and even in the large-$N$ limit some interesting physical properties
emerge.  It is therefore worth presenting a detailed discussion of
this system, which also offers the possibility of comparing the
different technical approaches to the large-$N$ solution in a
completely controlled situation.

The lattice formulation of the two-dimensional ${\rm U}(N)$ gauge
theory is based on dynamical variables $U_{x,\mu}$ which are defined on
links; however, because of gauge invariance, in two dimensions there
are no transverse gauge degrees of freedom, and a one-to-one
correspondence can be established between link variables and
plaquettes.  A convenient way of exploiting this fact consists in
fixing the gauge \cite{Gross-Witten}
\begin{equation}
U_{x,0} = 1
\label{temporal-gauge}
\end{equation}
(the lattice version of the temporal gauge $A_0 = 0$).  An extremely
important consequence of the gauge choice (\ref{temporal-gauge})
emerges from considering the gauge-fixed form of the single-plaquette
contribution to the lattice action:
\begin{equation}
\Tr\left(U_{x,0} U_{x+0,1} U_{x+1,0}^\dagger U_{x,1}^\dagger\right)
\to \Tr U_{x+0,1} U_{x,1}^\dagger .
\end{equation}
This is nothing but the single-link contribution to the
one-dimen\-sional lattice action of a principal chiral model whose
links lie along the 0 direction.  When considering invariant
expectation values (Wilson loops), we then recognize that they can be
reduced to contracted products of tensor correlations of variables
defined on decoupled one-dimensional models.  As a consequence, YM$_2$
factorizes completely into a product of independent chiral models
labeled by their 1 coordinate.  Not only the partition function, but
also all invariant correlations can be systematically mapped into
those of the corresponding chiral models.  The area law for non
self-interacting Wilson loops in YM$_2$ and the exponential decay of
the two-point correlations in one-dimensional chiral models are
trivial corollaries of these results \cite{Gross-Witten}.

The above considerations allow us to focus on the prototype model
defined by the action
\begin{equation}
S = - N \sum_i \Tr(U_i U_{i+1}^\dagger + U_i^\dagger U_{i+1}),
\label{S-proto}
\end{equation}
where $i$ is the site label of the one-dimensional lattice.  By
straightforward manipulations we may show that the most general
nontrivial correlation one really needs to compute involves product of
invariant operators of the form
\begin{equation}
\Tr(U_0 U_l^\dagger)^k,
\end{equation}
where $l$ plays the r\^ole of the space distance, and $k$ is a sort
of ``winding number''.

An almost trivial corollary of the above analysis is the observation
that YM$_2$ and principal chiral models in one dimension enjoy a
property of ``geometrization'', i.e., the only variables that can turn
out to be relevant for the complete determination of expectation
values are the single-plaquette (single-link) averages of products of
powers of moments \cite{Rossi-qcd2}
\begin{equation}
\prod_k \left[\Tr(U_0 U_1^\dagger)^k\right]^{m_k}
\label{mega-product}
\end{equation}
and the geometrical features of the correlations (in YM$_2$, areas of
Wilson loops and subloops; in chiral models, distances of correlated
points), such that all coupling dependence is incorporated in the
expectation values of the quantities (\ref{mega-product}).  This
result is sufficiently general to apply not only to the Wilson action
formulation, but also to all ``local'' actions such that the
interaction depends only on invariant functions of the
single-plaquette (single-link) variable, i.e., any linear combination
of the expressions appearing in Eqs.\ (\ref{S-spin}), (\ref{S-gauge})
\cite{Rossi-qcd2,Jurkiewicz-Zalewski,Chen-Tan-Zheng-universality}.

In order to proceed to the actual computation, it is convenient to
perform a change of variables, allowed by the invariance of the Haar
measure, parameterizing the fields by
\begin{equation}
V_l = U_{l-1} U_l^\dagger;
\label{chiral-variable-change}
\end{equation}
the action (\ref{S-proto}) explicitly factorizes into
\begin{equation}
S = -N \sum_l \Tr(V_l + V_l^\dagger).
\end{equation}
It is now easy to get convinced that in the most general case a Wilson
loop expectation value (correlation function) can be represented as a
finite product of invariant tensors, each of which is originated by a
single-link integration of the form
\begin{equation}
{\int\d V_l \, f(V_l) \exp\left[N \beta \Tr(V_l + V_l^\dagger)\right]
\over \int\d V_l \exp\left[N \beta\Tr(V_l + V_l^\dagger)\right]}
\equiv \left<f(V_l)\right>,
\end{equation}
where $f(V_l)$ is any (tensor) product of $V_l$'s and $V_l^\dagger$'s,
and the only nontrivial contributions to the full expectation value
come from integrations extended to plaquettes belonging to the area
enclosed by the loop itself (in chiral models, links comprised between
the extremal points of the space correlation).

For sake of definiteness, we may focus on the correlators
\cite{Rossi-Vicari-QCD2}
\begin{equation}
W_{l,k} \equiv {1 \over N}\left<\Tr(U_0 U_l^\dagger)^k\right>,
\end{equation}
and find that
\begin{eqnarray}
W_{l,k} = {\int\d V_1 ...\d V_l\,(1/N) \Tr(V_1 ... V_l)^k
    \exp\left[N \beta \sum_{i=1}^l \Tr(V_i + V_i^\dagger)\right]
\over \prod_i 
    \int\d V_i \exp\left[N \beta\Tr(V_i + V_i^\dagger)\right]} \, .
\nonumber \\
\end{eqnarray}
This problem can be formally solved for arbitrary $N$ by a character
expansion, which we shall discuss in Subs.\ \ref{character-expansion}.
It is however immediate to recognize that we are ultimately led to
computing the general class of group integrals whose form is
\begin{equation}
\int\d V \prod_k\left(\Tr V^k\right)^{\!\!m_k}
\exp\left[N \beta \Tr(V + V^\dagger)\right]
\label{general-group-integral}
\end{equation}
(where the product runs over positive and negative values of $k$), and
in turn it is in principle an exercise based on the exploitation of
the result for the external fiend single-link integral introduced in
Eq.\ (\ref{one-link}).

By the way, integrals of the form (\ref{general-group-integral}) can
easily be expressed as linear combinations of integrals belonging to
the class
\begin{equation}
\int\d V \, \chi_{(\lambda)}(V)
\exp\left[N \beta \Tr(V + V^\dagger)\right],
\label{character-goup-integral}
\end{equation}
where $\lambda$ labels properly chosen representations of ${\rm U}(N)$.
Eq.\ (\ref{character-goup-integral}) is in turn related to the
definition of the character coefficients in the character expansion of 
$\exp[N \beta \Tr(V + V^\dagger)]$.  For arbitrary $N$, as a matter of
principle, $\chi_{(\lambda)}(V)$ has a representation in terms of the
eigenvalues $\phi_i$ of the matrix $V$, while $\Tr(V + V^\dagger) =
2\sum_i \cos\phi_i$ and the measure itself can in this case be
expressed in terms of the eigenvalues as
\begin{equation}
\d\mu(V) \sim \prod_i \d\phi_i\,\Delta^2(\phi_1,...,\phi_N),
\end{equation}
where
\begin{eqnarray}
\Delta(\phi_1,...,\phi_N) &\equiv& \det\exp\Vert\I(i\phi_j)\Vert,
\nonumber \\ 
\Delta^2(\phi_1,...,\phi_N) &=& 
\prod_{i<j} 4\sin^2{\phi_i-\phi_j \over 2} \,.
\end{eqnarray}
As a consequence, it is always possible to express all ${\rm U}(N)$
integrals in the class (\ref{character-goup-integral}) in terms of
linear combinations of products of modified Bessel functions
$I_k(2N\beta)$, with $k<N$.

Let us now come to the specific issue of evaluating the relevant
physical quantities in the large-$N$ limit of ${\rm U}(N)$ models, and
comparing the procedures corresponding to different possible
approaches.  Basic to most subsequent developments is the observation
that the large-$N$ factorization property allows us to focus on a very
restricted class of interesting correlations, which we label by
\begin{equation}
w_k \equiv \left<{1 \over N} \Tr V^k\right> \equiv W_{1,k} \,.
\end{equation}

The first explicit solution to the problem of evaluating $w_k$ in the
large-$N$ limit was offered by Gross and Witten \cite{Gross-Witten}.
To this purpose, they introduced the eigenvalue density
\begin{equation}
\rho(\phi) = {1\over N} \sum_i \delta(\phi-\phi_i),
\end{equation}
and considered the group integral defining the partition function of
the single-link model
\begin{equation}
Z(\beta) \sim \int\prod_i\d\phi_i \, \Delta^2(\phi_1,...,\phi_N)
\exp\!\left(2 N \beta \sum_i\cos\phi_i\right).
\label{single-link-Z}
\end{equation}
The integral (\ref{single-link-Z}) can be evaluated in the
$N\to\infty$ limit by a saddle-point technique
\cite{Brezin-Itzykson-ZinnJustin-Zuber} applied to the
effective action 
\begin{equation}
2\beta \int \rho(\phi) \cos(\phi)\,\d\phi +
    \princint \rho(\phi)\,\rho(\phi') 
    \log\sin{\phi-\phi'\over2}\,\d\phi\,\d\phi' ,
\end{equation}
with the constraint $\int \rho(\phi)\,\d\phi = 1$.  The support of the
function $\rho(\phi)$ is dynamically determined.  The saddle-point
integral equation is
\begin{equation}
2\beta \sin\phi = \int_{-\phi_c}^{\phi_c}\d\phi' \,
    \rho(\phi') \cot{\phi-\phi'\over2} \,,
\end{equation}
and it is possible to identify two distinct solutions, corresponding
to weak and strong coupling.  When $\beta$ is small, it is easy to
find out that 
\begin{equation}
\rho(\phi) = {1\over2\pi} (1 + 2\beta\cos\phi), \qquad
-\pi \le \phi \le \pi ;
\label{easy-rho-strong-coupling-UN}
\end{equation}
$\rho(\phi)$ is positive definite whenever $\beta\le\casefr{1}{2}$.
When $\beta$ is large, $\phi_c<\pi$ and
\begin{equation}
\rho(\phi) = {2\beta\over\pi} \cos{\phi\over2}
    \sqrt{{1\over2\beta} - \sin^2{\phi\over2}}, \qquad
\sin^2{\phi_c\over2} = {1\over2\beta} \,,
\label{easy-rho-weak-coupling-UN}
\end{equation}
submitted to the condition $\beta\ge\casefr{1}{2}$.  Therefore it is
possible to identify the location of the third-order phase transition
\cite{Gross-Witten}:
\begin{equation}
\beta_c = \casefr{1}{2} \,.
\label{beta_c-UN}
\end{equation}
By direct substitution, one finds the values of the free and internal
energy (per unit link or unit plaquette):
\begin{equation}
{F \over N^2} = 
\left\{
\renewcommand\arraystretch{1.3}
\begin{array}{l@{\quad}l}
\beta^2 \,, & \beta\le\casefr{1}{2} \,, \\
2 \beta - \casefr{1}{2} \log 2\beta - \casefr{3}{4} \,, & 
\beta\ge\casefr{1}{2} \,,
\end{array}
\right.
\label{F-UN}
\end{equation}
\begin{equation}
w_1 = {1\over2}\,{\partial\over\partial\beta}\,{F \over N^2} = 
\left\{
\renewcommand\arraystretch{1.3}
\begin{array}{l@{\quad}l}
\beta \,, & \beta\le\casefr{1}{2} \,, \\
\displaystyle 1 - {1\over4\beta} \,, & \beta\ge\casefr{1}{2} \,.
\end{array}
\right.
\label{w1-UN}
\end{equation}
More generally, one may evaluate $w_k$ from $\rho(\phi)$, thanks to
the relationship
\begin{eqnarray}
w_k &=& \int_{-\phi_c}^{\phi_c}\d\phi \cos k\phi\,\rho(\phi) 
\nonumber \\ &=&
\left\{
\renewcommand\arraystretch{1.3}
\begin{array}{l@{\quad}l}
0 \,, & \beta\le\casefr{1}{2},\ k \ge 2 \,, \\
\displaystyle \left(1 - {1\over2\beta}\right)^{\!\!2}
{1\over k-1}\,P^{(1,2)}_{k-2}\!\left(1 - {1\over\beta}\right), &
\beta\ge\casefr{1}{2} \,, 
\end{array}
\right.
\end{eqnarray}
where $P^{(\alpha,\beta)}_k$ are the Jacobi polynomials.  All $w_k$
are differentiable once in $\beta=\beta_c$, but their second
derivatives are discontinuous.  Let us notice that
Eqs.\ (\ref{beta_c-UN}), (\ref{F-UN}), and (\ref{w1-UN}) are an
immediate consequence of Eqs.\ (\ref{SD-Z-strong-cond}) and
(\ref{SD-W}) for the special choice
\begin{equation}
x_s = \beta^2 \,.
\end{equation}

\subsection{The Schwinger-Dyson equations of the two-dimensional
Yang-Mills theory}
\label{sd-YM}

It is interesting to obtain the above results from the algebraic
approach to the Schwinger-Dyson equations of the model.  We can
restrict Eqs.\ (\ref{Migdal-Makeenko}) to the set of Wilson loops
${\cal C}_k$ consisting of $k$ turns around a single plaquette, in
which case by definition $W({\cal C}_k) = w_k$.  Formally, the
Schwinger-Dyson equations do not close on this set of expectation
values; however, one may check by inspection, using the factorization
property of two-dimensional functional integral for the Yang-Mills
theory, that contributions from other Wilson loops cancel in the
equations for $w_k$ (this is strictly a two-dimensional property).  As
a consequence, we obtain the large-$N$ relationships
\cite{Paffuti-Rossi-solution}
\begin{equation}
\beta(w_{n-1} - w_{n+1}) = \sum_{k=1}^n w_k w_{n-k} \,,
\label{MM-reduced}
\end{equation}
with a boundary condition $w_0 = 1$.  The solution is found by
defining a generating function
\begin{equation}
\Phi(t) \equiv \sum_{k=0}^\infty w_k t^k
\end{equation}
and noticing that Eq.\ (\ref{MM-reduced}) corresponds to
\begin{equation}
\Phi t^2 - (\Phi - 1 - w_1 t) = 
{t\over\beta} (\Phi^2 - \Phi),
\end{equation}
which is solved by
\begin{eqnarray}
\Phi(t) = {\beta\over2t} 
\sqrt{\left(1 + {t\over\beta} + t^2\right)^{\!\!2}
    - 4 t^2 \left(1 - {w_1\over\beta}\right)}
- {\beta\over2t} \left(1 - {t\over\beta} - t^2\right).
\nonumber \\
\label{Phi-strong-coupling}
\end{eqnarray}
The condition $|w_k|\le1$ implies that $\Phi(t)$ is holomorphic within
the unitary circle.  On the boundary of the analyticity domain, $t =
\e^{\I\phi}$ and
\begin{equation}
w_k = {1\over\pi} \int_{-\pi}^\pi 
    [\Re\Phi(\phi) - \casefr{1}{2}] \cos k\phi \, \d\phi,
\end{equation}
and as a consequence we may identify
\begin{equation}
\Re\Phi(\phi) - \casefr{1}{2} = \rho(\phi).
\end{equation}
The positivity condition on $\rho(\phi)$ leads to a complete
determination of the solution, implying either
\begin{equation}
w_1 = \beta, \qquad w_k = 0 \ \ (k\ge2), \qquad \beta\le\casefr{1}{2}
\end{equation}
or $\rho(\pi) = 0$, which in turn leads to
\begin{equation}
w_1 = 1 - {1\over4\beta}, \qquad -\phi_c\le\phi\ge\phi_c, 
\qquad \beta\le\casefr{1}{2} \, ,
\end{equation}
and $\phi_c$ is given by Eq.\ (\ref{easy-rho-weak-coupling-UN}).
It is immediate to check that the resulting eigenvalue densities are
the same as Eqs.\ (\ref{easy-rho-strong-coupling-UN}) and
(\ref{easy-rho-weak-coupling-UN}).

Let us mention that these methods may in principle be applied to more
general formulation of the theory based on ``local'' actions, and in
particular Wilson loop expectation values can be computed for the
fixed-point version of the model, corresponding to the continuum
action \cite{Rossi-qcd2}.  The fixed-point action in YM$_2$ in turn is
nothing but the ``heat kernel'' action \cite{Drouffe-heat}, discussed
in the large-$N$ context in Ref.\ \cite{Menotti-Onofri}.  Large-$N$
continuum YM$_2$ is slightly beyond the purpose of the present review.
We must however mention that in recent years a number of interesting
results have appeared in a string theory context.  It is worth quoting
Refs.\ \cite{Rusakov,Douglas-Kazakov,Gross-Taylor} and references
therein.

While the problem of evaluating the more general expectation values
$W_{l,k}$ is solved in principle, in practice it is not always simple
to obtain compact closed-form expressions whose general features can
be easily understood.  In the strong-coupling regime
$\beta<\casefr{1}{2}$, it is not too difficult to determine from
finite-$N$ results the large-$N$ limit in the form
\cite{Rossi-Vicari-QCD2}
\begin{equation}
\lim_{N\to\infty} W_{l,k} = {(-1)^{k-1}\over k}
\left(
\renewcommand\arraystretch{1.3}
\begin{array}{c}
lk - 2 \\
k - 1
\end{array}
\right)
\beta^{kl},
\label{w_lk-SU}
\end{equation}
and one may show that the corresponding Schwinger-Dyson equations
close on the set $W_{l,k}$ for any fixed $l$ and are solved by Eq.\
(\ref{w_lk-SU}).  As a matter of fact, by defining
\begin{equation}
\Phi_l(t) \equiv \sum_{k=0}^\infty W_{l,k} t^k,
\end{equation}
one may show that the strong-coupling Schwinger-Dyson equations reduce
to
\begin{equation}
[\Phi_l(t) - 1] [\Phi_l(t)]^{l-1} = \beta^l t .
\end{equation}
For the interesting values $l=1$ and $l=2$, Eq.\ (\ref{w_lk-SU})
reduces to
\begin{equation}
\Phi_1(t) = 1 + \beta t,
\end{equation}
consistent with the strong-coupling solution
(\ref{Phi-strong-coupling}), and
\begin{equation}
\Phi_2(t) = \casefr{1}{2}\left(\sqrt{1 + 4 \beta^2 t^2} + 1\right),
\end{equation}
related to the generating function for the moments of the energy
density
\begin{eqnarray}
&&{1 \over N} 
\left<\Tr{1 \over 1 - \beta t (V_n + V_{n+1}^\dagger)}\right>
= 1 + 2 t \beta^2 + 2 \sum_{k=1}^\infty (\beta t)^{2k} W_{2,k} 
\nonumber \\
&&\qquad=\, 2 \beta^2 t + \sqrt{1 + 4 \beta^4 t^2} \,.
\label{moments-generating-function-UN}
\end{eqnarray}

Eq.\ (\ref{moments-generating-function-UN}) is related to a different
approach for solving large-$N$ unitary-matrix models, based on an
integration of the matrix angular degrees of freedom to be performed
in strong coupling \cite{Kazakov-Kozhamkulov-Migdal,Barsanti-Rossi}.

The corresponding weak-coupling problem is definitely more difficult.
As far as we can see, the Schwinger-Dyson equations close only on a
larger set of correlation functions, defined by the generating
function \cite{Rossi-unpublished}
\begin{eqnarray}
&&D_{k,n}^{(l)}(t) =
{1 \over N} \Tr \left[(V_k)^{n+1} V_{k+1} ... V_l \,
{1 \over 1 - t V_1 ... V_l}\right], \nonumber \\
&&0 < k \le l,\ n \ge 0,
\end{eqnarray}
such that
\begin{equation}
\Phi_l(t) = 1 + t D_{1,0}^{(l)}(t).
\end{equation}
The explicit form of the equations is
\begin{eqnarray}
&&\sum_{j=0}^{n-1} w_j D_{k,n-j}^{(l)}(t) +  
    D_{l,n-1}^{(l)}(t)\, D_{k,0}^{(l)}(t) \nonumber \\
&&\quad+\,
    \beta\left[D_{k,n+1}^{(l)}(t) - D_{k,n-1}^{(l)}(t)\right] = 0,
\qquad 1 \le k \le l .
\end{eqnarray}
When $l=1,2$ it is possible to find explicit weak-coupling solutions,
but the general case $l>2$ has not been solved so far.

More about the calculability of Wilson loops with arbitrary contour in
two-dimensional ${\rm U}(\infty)$ lattice gauge theory can be found in
Ref.\ \cite{Kazakov-Kostov-wilson}.  The corresponding continuum
calculations are presented for arbitrary ${\rm U}(N)$ groups in
Ref.\ \cite{Kazakov-wilson}.

\subsection{Large-$N$ properties of the determinant}
\label{determinant}

It is quite interesting to apply the results of Subs.\ 
\ref{properties-determinant}, concerning the properties of the
determinant, to YM$_2$ and principal chiral models in one dimension.
Exploiting the factorization of the functional integration and the
possibility of performing the variable change
(\ref{chiral-variable-change}) in the operators as well as in the
action, we can easily obtain the relationship
\begin{equation}
\Delta_l \equiv \det\left[U_0 U^\dagger_l\right] =
\det\left[V_1 ... V_l\right] = \det V_1 ... \det V_l,
\end{equation}
and, as a consequence,
\begin{equation}
\left<\Delta_l\right> = \left<\det V_1\right>^l.
\end{equation}
The problem is therefore reduced to that of evaluating $\left<\det
V\right>$ in the single-plaquette model.  It is immediate to recognize
from Eqs.\ (\ref{SD-Xl-reduced}) and (\ref{SD-Xl-sol}) that
\begin{eqnarray}
&\displaystyle \left<\det V\right> \to \sqrt{1 - {1\over2\beta}}, 
&\qquad \beta \ge \casefr{1}{2}, 
\label{detV-weak-coupling} \\
&\displaystyle \left<\det V\right> \to 0, 
&\qquad \beta \le \casefr{1}{2}.
\end{eqnarray}
Apparently, this expectation value acts as an order parameter for the
phase transition between the weak- and strong-coupling phases.  More
precisely, according to Green and Samuel
\cite{Green-Samuel-un1,Green-Samuel-largeN}, one must identify the
order parameter with the quantity
\begin{equation}
\left<\Delta_l\right>^{1/N}
\end{equation}
and notice that
\begin{eqnarray}
&\displaystyle \left<\Delta_l\right>^{1/N} \to 1 
&\qquad \hbox{in weak coupling,} 
\label{Deltal-wc} \\
&\displaystyle \left<\Delta_l\right>^{1/N} \to \exp(-\sigma l) 
&\qquad \hbox{in strong coupling,} 
\label{Deltal-sc}
\end{eqnarray}
where $\sigma$ acts as a ${\rm U}(1)$ ``string tension''.  Eqs.\
(\ref{Deltal-wc}) and (\ref{Deltal-sc}) generalize to higher
dimensions, when replacing $l$ with the (large) area of the
corresponding Wilson loop.  Notice that the weak-coupling result is
consistent with the decoupling of the ${\rm U}(1)$ degrees of freedom
from the ${\rm SU}(N)$ degrees of freedom, and with the interpretation
of ${\rm U}(1)$ as a free massless field.

It is therefore interesting to compute 
\begin{equation}
\sigma = -{1\over N} \log\left<\det V\right>
\end{equation}
in the case of the single-matrix model; this requires taking the
large-$N$ limit only after the strong-coupling calculation of
$\left<\det V\right>$ has been performed.  Since the technique of
evaluation of $\sigma$ has some relevance for subsequent developments,
we shall briefly sketch its essential steps.  Standard manipulations
of the single-link integrals for finite $N$ allow to evaluate
\begin{eqnarray}
A_{m,N}(\beta) &=& \int\d V \exp\left[N \beta\Tr(V+V^\dagger)\right] 
  (\det V)^m = \det\Vert I_{k-l-m}(2N\beta)\Vert. \nonumber \\
\end{eqnarray}
These quantities can be shown to satisfy the recurrence relations
\cite{Guha-Lee-chiral}
\begin{equation}
A^2_{m,N} - A_{m+1,N} A_{m-1,N} = A_{m,N-1} A_{m,N+1} .
\label{A-recurrence}
\end{equation}
Willing to compute expectation values, we define
\begin{equation}
\Delta_{m,N}(\beta) = \left<(\det V)^m\right> = {A_{m,N}\over A_{0,N}}
\,.
\end{equation}
Eq.\ (\ref{A-recurrence}) implies that
\begin{equation}
\Delta^2_{m,N} - \Delta_{m+1,N} \Delta_{m-1,N} = 
\Delta_{m,N-1} \Delta_{m,N+1} (1 - \Delta^2_{1,N}).
\label{Delta-recurrence}
\end{equation}

Since all $\Delta_{m,1}$ are known, it is possible to reconstruct all
$\Delta_{m,N}$ from Eq.\ (\ref{Delta-recurrence}) once $\Delta_{1,N}$
is determined.  Now $\Delta_{1,N}$ is exactly $\left<\det V\right>$,
and it is possible to show that it obeys the following second-order
differential equation  \cite{Rossi-exact}
\begin{eqnarray}
&&{1\over s}\,{\d\over\d s}\,s\,{\d\over\d s}\,\Delta_{1,N} +
{1\over1-\Delta^2_{1,N}}\left[\left({\d\over\d s}
        \,\Delta_{1,N}\right)^{\!\!2}
    - {N^2\over s^2}\right]\Delta_{1,N} \nonumber \\
&&\quad+\,
    (1-\Delta^2_{1,N})\Delta_{1,N} = 0,
\label{Delta1N-differential}
\end{eqnarray}
where $s = 2N\beta$.  Eq.\ (\ref{Delta1N-differential}) can be analyzed
in weak and strong coupling and in the large-$N$ limit.  In particular
the weak-coupling $1/N$ expansion leads to
\begin{equation}
\Delta_{1,N} \to \sqrt{1 - {1\over2\beta}} - {1\over N^2}\,
    {1\over128\beta^3}\left(1 - {1\over2\beta}\right)^{\!\!5/2}
    + O\left(1\over N^4\right),
\end{equation}
thus confirming Eq.\ (\ref{detV-weak-coupling}), while in strong
coupling one may show that
\begin{equation}
\Delta_{1,N} = J_N(2N\beta) + O(\beta^{3N+2}) 
\goto_{N\to\infty} J_N(2N\beta),
\end{equation}
where $J_N$ is the standard Bessel function, whose asymptotic behavior
is well known.  As an immediate consequence, we find
\begin{equation}
-\sigma = \sqrt{1-4\beta^2} - \log{1 + \sqrt{1-4\beta^2} 
        \over 2\beta}, \qquad \beta<\casefr{1}{2}. 
\end{equation}
This result was first guessed by Green and Samuel
\cite{Green-Samuel-largeN}, and then explicitly demonstrated in
Ref.\ \cite{Rossi-exact}.

\subsection{Local symmetry breaking in the large-$N$ limit}
\label{local-SB}

Another interesting application of the external-field single-link
integral to the large-$N$ limit of two-dimensional Yang-Mills theories
is the study of the possibility of breaking a local symmetry, as a
consequence of the thermodynamical nature of the limit.  If we
introduce an infinitesimal explicit ${\rm U}(N)$ symmetry breaking
term in the action \cite{Celmaster-Green}
\begin{equation}
S = - \beta N \left[\Tr V + J N V^{ij} + \hbox{h.c.}\right],
\end{equation}
corresponding to replacing
\begin{equation}
A_{lm} \to \beta\left[\delta_{lm} + N J \delta_{lj}\delta_{mi}\right]
\end{equation}
in Eq.\ (\ref{one-link}), we find that the eigenvalues of $AA^\dagger$
are 
\begin{eqnarray}
x_{1,2} &=& \beta^2 \left[1 + \casefr{1}{2} N^2 J^2 \pm \casefr{1}{2} 
    \sqrt{J^4 N^4 + 4 J^2 N^2}\right], \nonumber \\
x_l &=& \beta^2, \qquad l>2 . 
\end{eqnarray}
When taking the large-$N$ limit of the free energy, we find
\begin{equation}
\lim_{N\to\infty} {\log Z\over N^2} = F_0(\beta) + 2 \beta|J|,
\end{equation}
and in the limit $J\to0^\pm$ we then find
\begin{equation}
\left<\Re V^{ij}\right> = \pm 1. 
\end{equation}
We therefore expect that, for finite $k$, the ${\rm U}(k)$ global
symmetries of large-$N$ chiral models and ${\rm U}(k)$ gauge
symmetries are broken in any number of dimensions
\cite{Celmaster-Green}.  This phenomenon cannot occur for any finite
value of $N$ in two dimensions.

\subsection{Evaluation of higher-order corrections}
\label{higher-order-corrections}

In the context of large-$N$ two-dimensional Yang-Mills theory, it is
worth mentioning that it is possible to compute systematically
higher-order corrections to physical quantities in the powers of
$1/N^2$.  It is interesting to notice that the weak-coupling
corrections to the free energy \cite{Goldschmidt} (see also
\cite{Muller-Ruhl-mean-1})
\begin{eqnarray}
F &=& F_0 + {1\over N^2}\left[{1\over12} - A - {1\over12} \log N 
        - {1\over8} \log\left(1 - {1\over2\beta}\right)\right] 
\nonumber \\ &+&\,
    {1\over N^4}\left[{3\over1024\beta^3}
        \left(1 - {1\over2\beta}\right)^{\!\!-3} 
  - {1\over240}\right] + ... \,, \\
U &=& 1 - {1\over4\beta} - 
    {1\over N^2}\,{1\over32\beta^2} 
        \left(1 - {1\over2\beta}\right)^{\!\!-1}
\nonumber \\ &-&\,
    {1\over N^2}\,{1\over1024\beta^4} 
        \left(1 - {1\over2\beta}\right)^{\!\!-4} +
    O\left(1\over N^4\right), 
\end{eqnarray}
where $A = 0.24875...$, are well defined, but become singular when
$\beta\to\casefr{1}{2}$.  In turn, when evaluating higher-order
corrections in the strong-coupling phase, one finds out that there are
no corrections proportional to powers of $1/N$, while there are
contributions that fall off exponentially with large $N$, as expected
from the general arguments discussed in
Subs.\ \ref{mean-field+strong-coupling} in connection with the
appearance of the DeWit-'t Hooft poles.

Let us however mention that Eqs.\ (\ref{Delta-recurrence}) and
(\ref{Delta1N-differential}) are also the starting point for a
systematic $1/N$ expansion of the free energy in the weak-coupling
regime, alternative to Goldschmidt's procedure.  The basic ingredient
is the observation that, defining the free energy at finite $N$ by
\begin{equation}
F_N(\beta) = \log A_{0,N}(\beta),
\end{equation}
one may show that
\begin{equation}
{\d\over\d s}(\log F_N - \log F_{N-1}) =
{\Delta_{1,N}\over1-\Delta_{1,N}^2} 
    \left({\d\over\d s}\,\Delta_{1,N} + 
        {N\over s}\,\Delta_{1,N}\right),
\end{equation}
and this allows for a systematic reconstruction of $F_N$, whose
strong-coupling form is \cite{Guha-Lee-chiral}
\begin{equation}
F_N(\beta) = N^2\beta^2 - \sum_{k=1}^\infty k J_{N+k}^2(2N\beta)
+ O(\beta^{4N+4}).
\end{equation}

\subsection{Mixed-action models for lattice YM$_2$}
\label{mixed-ation}

Another instance of the problem of the single-link integration for
matrix fields in the adjoint representation of the full symmetry group
occurs in the discussion of the so-called ``mixed action'' models.
Consider the following single-link integral
\cite{Chen-Tan-Zheng-phase}, resulting from a different formulation of
lattice YM$_2$,
\begin{equation}
Z(\beta_{\rm f},\beta_{\rm a}) = \int\d U
\exp\left\{N\beta_{\rm f}\Tr(U+U^\dagger) 
    + \beta_{\rm a} |\Tr U|^2\right\}.
\label{Z-fa}
\end{equation}
It is possible to show that, in the large-$N$ limit, the corresponding
free energy can be obtained by the same saddle-point technique
presented in Subs.\ \ref{2d-YM-single-link}, i.e., by introducing a
spectral density $\rho(\theta)$ for the eigenvalues of $U$.  This
spectral density turns out to be precisely the same as the one
obtained when $\beta_{\rm a} = 0$, if one simply replaces $\beta_{\rm
f}$ by an effective coupling
\begin{equation}
\beta_{\rm eff} = \beta_{\rm f} + \beta_{\rm a} w_1(\beta_{\rm eff}),
\label{beff-def}
\end{equation}
where $w_1$ can be evaluated in terms of $\rho(\theta)$ as
\begin{equation}
w_1(\beta_{\rm eff}) = \int\d\theta \cos\theta\,\rho(\theta).
\label{w1-eff}
\end{equation}
Eq.\ (\ref{w1-eff}) is a self-consistency condition for $w_1$, which
allows a determination of 
$\beta_{\rm eff}(\beta_{\rm f},\beta_{\rm a})$.
Finally, by substitution into the effective action, one finds the
relationship 
\begin{equation}
F(\beta_{\rm f},\beta_{\rm a}) = 
F(\beta_{\rm eff}(\beta_{\rm f},\beta_{\rm a}),0) -
\beta_{\rm a} w_1^2(\beta_{\rm eff}(\beta_{\rm f},\beta_{\rm a})),
\label{F-fa}
\end{equation}
where $F(\beta,0)$ is nothing but the free energy obtained in
Subs.\ \ref{2d-YM-single-link}. 

The strong- and weak-coupling solutions are separated by the line 
$2\beta_{\rm f} + \beta_{\rm a} = 1.$  In strong coupling one obtains
\begin{eqnarray}
\beta_{\rm eff} &=& w_1 = {\beta_{\rm f}\over1-\beta_{\rm a}}\,,
\nonumber \\
F &=& {\beta_{\rm f}^2\over1-\beta_{\rm a}}\,,
\end{eqnarray}
while in weak coupling
\begin{eqnarray}
\beta_{\rm eff} &=& {1\over2} \left[\beta_{\rm f} + \beta_{\rm a} +
    \sqrt{(\beta_{\rm f} + \beta_{\rm a})^2 - \beta_{\rm a}}\right],
\nonumber \\
w_1 &=& {1\over2\beta_{\rm a}} \left[\beta_{\rm a} - \beta_{\rm f} +
    \sqrt{(\beta_{\rm f} + \beta_{\rm a})^2 - \beta_{\rm a}}\right],
\nonumber \\
F &=& \beta_{\rm f} + {\beta_{\rm a}\over2} 
    - {\beta_{\rm f}^2\over2\beta_{\rm a}} - {1\over2} 
    - {1\over2}\log\left[\beta_{\rm f} + \beta_{\rm a} +
    \sqrt{(\beta_{\rm f} + \beta_{\rm a})^2 - \beta_{\rm a}}\right]
\nonumber \\
  &&\quad+\,
    {1\over2}\left(1 + {\beta_{\rm f}\over\beta_{\rm a}}\right)
    \sqrt{(\beta_{\rm f} + \beta_{\rm a})^2 - \beta_{\rm a}}.
\end{eqnarray}
It may be interesting to quote explicitly the limiting case 
$\beta_{\rm f} = 0$, where \cite{Brihaye-Rossi-weak}
\begin{eqnarray}
Z(0,\beta_{\rm a}) &\equiv& \int\d U \exp\beta_{\rm a}|\Tr U|^2 
\nonumber \\
&=& \left\{
\renewcommand\arraystretch{1.3}
\begin{array}{l@{\quad}c}
0, & \beta_{\rm a} < 1, \\
\displaystyle {1\over2}\beta_{\rm a} + 
{1\over2}\beta_{\rm a}\sqrt{1 - {1\over\beta_{\rm a}}} -
{1\over2}\log\beta_{\rm a}
    \left(1 + \sqrt{1 - {1\over\beta_{\rm a}}}\right), &
\beta_{\rm a} > 1.
\end{array}
\right. \nonumber \\
\end{eqnarray}

One may actually show that, in any number of dimensions, a lattice
gauge theory with mixed action
\cite{Samuel-adjoint,Makeenko-Polikarpov,Samuel-phase} (a trivial
generalization of Eq.\ (\ref{Z-fa})) is solved in the large-$N$ limit
in terms of the solution of the corresponding theory with pure Wilson
action; Eqs.\ (\ref{beff-def}) and (\ref{F-fa}) hold as they stand, and
\begin{equation}
w_1(\beta_{\rm eff}) = \left.{1\over N} \left<\Tr U_p\right>
\right|_{\beta_{\rm f} = \beta_{\rm eff},\ \beta_{\rm a} = 0} .
\end{equation}
More about the large-$N$ behavior of variant actions can be found in
Refs.\ \cite{Ogilvie-Horowitz,Jurkiewicz-KorthalsAltes,%
Jurkiewicz-KorthalsAltes-Dash}.  Different kinds of variant actions
have been studied in the large-$N$ limit in Refs.\
\cite{Rodrigues-variant,Lang-Salomonson-Skagerstam-third,Samuel-heat}.

\subsection{Double-scaling limit of the single-link integral}
\label{double-scaling-single-link}

In the Introduction, we mentioned that one of the most interesting
phenomena related to the large-$N$ limit of matrix models is the
appearance of the so-called ``double-scaling limit'' 
\begin{equation}
\left\{
\renewcommand\arraystretch{1.3}
\begin{array}{l}
N\to\infty, \\
g\to g_c,
\end{array}
\right.
\qquad N^{2/\gamma_1}(g_c-g) = \hbox{const}, 
\end{equation}
where $g$ is a (weak) coupling related to the inverse of $\beta$.  We
already discussed the general physical interpretation of this limit as
an alternative description of two-dimensional quantum gravity and its
relationship to the theory of random surfaces.  Here we only want to
consider the double-scaling limit properties for those simple models
of unitary matrices that can be reformulated as a single-link model
(cfr.\ Ref.\ \cite{Demeterfi-Tan}).

This specific subject was pioneered by Periwal and Shevitz
\cite{Periwal-Shevitz}, who discussed the double-scaling limit in
models belonging to the class
\begin{equation}
Z_N = \int\d U \exp\left[N\beta\Tr {\cal V}(U+U^\dagger)\right],
\label{ZN-ds}
\end{equation}
where ${\cal V}(U)$ is a polynomial in $U$.  Because of the invariance
of the measure, Eq.\ (\ref{ZN-ds}) can be reduced to 
\begin{equation}
Z_N \sim \int\d\phi_i |\Delta(\e^{\I\phi_1},...,\e^{\I\phi_N})|^2
 \exp\left[N\beta \textstyle\sum_i {\cal V}(2\cos\phi_i)\right],
\end{equation}
and solved by the method of orthogonal polynomials.  One starts by
defining polynomials 
\begin{equation}
P_n(z) = z^n + \sum_{k=0}^{n-1} a_{k,n} z^k,
\end{equation}
that satisfy
\begin{equation}
\oint {\d z\over2\pi\I z}\,P_n(z)\,P_m\!\left(1\over z\right)
\exp\left[N\beta {\cal V}\!\left(z + {1\over z}\right)\right] =
h_n\,\delta_{mn}\,,
\end{equation}
where the integration runs over the unit circle, and moreover obey the
recursion relation
\begin{equation}
P_{n+1}(z) = zP_n(z) + R_nz^nP_n\left(1\over z\right), \qquad
{h_{n+1}\over h_n} = 1 - R^2_n \,.
\end{equation}
where $R_n \equiv a_{0,n+1}$.  As a corollary,
\begin{equation}
Z_N \propto N! \prod_i\left(1-R^2_{i-1}\right)^{N-i},
\end{equation}
and one may show that
\begin{eqnarray}
(n+1)(h_{n+1}-h_n) &=& \oint {\d z\over2\pi\I z}
\exp\left[N\beta {\cal V}\!\left(z + {1\over z}\right)\right]
N\beta {\cal V}'\!\left(z + {1\over z}\right)
\nonumber \\ &\times&\,
\left(1 - {1\over z^2}\right)P_{n+1}(z)\,P_n\!\left(1\over z\right),
\end{eqnarray}
which in turn leads to a nonlinear functional equation for $R_n$.

The simplest example, corresponding to YM$_2$, amounts to choosing
${\cal V}' = 1$, obtaining
\begin{equation}
(n+1)R^2_n = N\beta R_n(R_{n+1}+R_{n-1})(1-R_n^2),
\label{YM2-ds}
\end{equation}
and in the large-$N$ limit, setting $n=N$ and $R_N=R$, we obtain the
limiting form
\begin{equation}
R^2 = 2\beta R^2(1-R^2),
\end{equation}
showing that $\beta_c=\casefr{1}{2}$ (degeneracy of solution $R_c=0$).
One may now look for the scaling solution to Eq.\ (\ref{YM2-ds}) in the
form
\begin{equation}
R_N-R_c = R_N = N^{-\mu} f\left[N^\rho(g_c-g)\right], \qquad 
g = {1\over\beta},
\end{equation}
where $f^2$ is related to the second derivative of the free energy.
This is a consistent Ansatz when 
\begin{equation}
\mu=\casefr{1}{3}, \qquad
\rho=\casefr{2}{3},
\label{k=1-exponents}
\end{equation}
leading to the equation
\begin{equation}
-2xf + 2f^3 = f'', \qquad x=N^\rho(g_c-g).
\end{equation}

In the case ${\cal V}' = 1 + \lambda u$, one finds the equation
\begin{equation}
{1\over\beta} = -2(1-R^2)(-1-\lambda+3\lambda R^2),
\end{equation}
which reduces to $1/\beta = \casefr{3}{2}(1-R^4)$ when
$\lambda=\casefr{1}{4}$.  A scaling solution to the corresponding
difference equation requires $\mu=\casefr{1}{5}$ and
$\rho=\casefr{4}{5}$.
When ${\cal V}' = 1 + \lambda_1 u + \lambda_2 u^2$, multicriticality
sets at $\lambda_1=-\casefr{3}{7}$ and $\lambda_2=\casefr{1}{14}$, and
$1/\beta = \casefr{10}{7}(1-R^6)$, leading to the exponents
$\mu=\casefr{1}{7}$ and $\rho=\casefr{6}{7}$.
Rather general results can be obtained for an arbitrary order $k$ of
the polynomial ${\cal V}$: $\mu = 1/(2k+1)$, $\rho = 2k/(2k+1)$, and
$c = 1 - 6/(k(k+1))$.

The double-scaling limit can also be studied in the case of the
external-field single-link integral \cite{Gross-Newman}, and it was
found that its critical behavior is simple enough to be identified
with that of the $k=1$ unitary-matrix model.  In the language of
quantum gravity, the only effect of introducing $N^2$ real parameters
$A_{ij}$ is that of renormalizing the cosmological constant, without
changing the universality class of the critical point.

A few interesting features of the double-scaling limit for the $k=1$
model are worth a more detailed discussion \cite{Damgaard-Heller}.  In
particular let us recall that, according to Eq.\ (\ref{k=1-exponents}),
\begin{equation}
\rho = {2\over\gamma_1} = {2\over3},
\end{equation}
and therefore $\gamma_1=3$, implying $c=-2$.  We may now reinterpret
the double-scaling limit of matrix models as a finite-size scaling
with respect to the ``volume'' parameter $N$ in a two-dimensional
$N{\times}N$ space.  As a consequence, we obtain relationships with
more conventional critical exponents through the identification
$\gamma_1=2\nu$, which in turn by hyperscaling leads to a
determination of the specific heat exponent $\alpha = 2(1-\nu)$.
Numerically we obtain $\nu=\casefr{3}{2}$ and $\alpha=-1$.  The result
$\alpha=-1$ can be easily tested on the solution of the model
\begin{equation}
C(\beta) = {1\over2}\,\beta^2\,{\d^2F\over\d\beta^2} =
\left\{
\renewcommand\arraystretch{1.3}
\begin{array}{l@{\quad}l}
\beta^2,     & \beta\le\beta_c, \\
\casefr{1}{4}, & \beta\ge\beta_c,
\end{array}
\right.
\end{equation}
with $\beta_c=\casefr{1}{2}$,
consistent with a negative critical exponent $\alpha=-1$.

It is also interesting to find tests for the exponent $\nu$,
especially in view of the fact that the most direct checks are not
possible in absence of a proper definition for the relevant
correlation length.  Numerical studies have been performed by
considering the partition function zero $\beta_0$ closest to the
transition point $\beta_c=\casefr{1}{2}$, finding that the relationship
\begin{equation}
\Im\beta_0 \propto N^{-1/\nu}
\end{equation}
is rather well satisfied even for very low values of $N$; at $N\ge5$,
it is valid within one per mille.  Another test concerns the location
of the peak in the specific heat in ${\rm U}(N)$ models, whose
position $\beta_{\rm peak}(N)$ should approach $\beta_c$ with
increasing $N$.  Finite-size scaling arguments predict
\begin{equation}
\beta_{\rm peak}(N) \cong \beta_c + a N^{-1/\nu} ,
\end{equation}
and large-$N$ results are very well fitted by the choice
$\nu=\casefr{3}{2}$, $a\cong0.60$ 
\cite{Campostrini-Rossi-Vicari-chiral-3}.

\subsection{The character expansion and its large-$N$ limit: 
${\rm SU}(N)$ vs.\ ${\rm U}(N)$} 
\label{character-expansion}

The general features of the character expansion for lattice spin and
gauge models have been extensively discussed by different authors.  In
particular, Ref.\ \cite{Drouffe-Zuber}, besides offering a general
presentation of the issues, presents tables of character coefficients
for many interesting groups, including ${\rm U}(\infty) \cong {\rm
SU}(\infty)$, for the Wilson action.  Let us therefore only briefly
recall the fundamental points of this approach, which is relevant
especially in the analysis of the strong-coupling phase and of the
phase transition.

In Sect.\ \ref{unitary-matrices} we classified the representations and
characters of ${\rm U}(N)$ groups.  Because of the orthogonality and
completeness relations, every invariant function of $V$ can be
decomposed in a generalized Fourier series in the characters of $V$.
Let us now consider for sake of definiteness chiral models with action
given by Eq.\ (\ref{action-spin}); extension to lattice gauge theories
is essentially straightforward, at least on a formal level.
We can replace the Boltzmann factor corresponding to each lattice link
by its character expansion:
\begin{eqnarray}
&& \exp\left\{\beta N\Tr\left[U_x U^\dagger_{x+\mu} + 
    U_{x+\mu} U^\dagger_{x}\right]\right\} \nonumber \\
&=&
\exp \Biggl\{ N^2 F(\beta) \sum_{(r)} d_{(r)} \tilde z_{(r)}(\beta)
    \, \chi_{(r)}\bigl(U_x U^\dagger_{x+\mu}\bigr)\Biggr\},
\label{link-char-exp}
\end{eqnarray}
where the sum runs over all the irreducible representations of ${\rm
U}(N)$, $F(\beta)$ is the free energy of the single-link model
\begin{eqnarray}
F(\beta) = {1\over N^2}
    \log\int \d V \exp\left[N\beta\Tr(V+V^\dagger)\right] =
{1\over N^2}\,\log\det\Vert I_{j-i}(2N\beta)\Vert,
\nonumber \\
\label{single-link-F}
\end{eqnarray}
and $\tilde z_{(r)}(\beta)$ are the character coefficients, defined by
orthogonality and representable in terms of single-link integrals as
\begin{equation}
d_{(r)} \tilde z_{(r)}(\beta) = \left<\chi_{(r)}(V)\right> =
{\det\Vert I_{\lambda_i+j-i}(2N\beta)\Vert
\over \det\Vert I_{j-i}(2N\beta)\Vert} \,,
\end{equation}
with $\lambda$ defined by Eq.\ (\ref{lambda}).  We may notice that, for
any finite $N$, $\tilde z_{(r)}(\beta)$ are meromorphic functions of
$\beta$, with no poles on the real axis, which is relevant to the
series analysis.  However, singularities may develop, as usual, in the
large-$N$ limit.  Eqs.\ (\ref{link-char-exp}) and (\ref{single-link-F})
become rapidly useless with growing $N$.  However, an extreme
simplification occurs in the large-$N$ limit, owing to the property
\begin{eqnarray}
d_{(l,m)} \tilde z_{(l,m)}(\beta) =
{1\over n_+!}\,{1\over n_-!}\,\sigma_{(l)}\sigma_{(m)}\,
(N\beta)^{n_++n_-} \left[1 + O(\beta^{2N})\right],
\nonumber \\
\label{dz}
\end{eqnarray}
where $n_+ = \sum_i l_i$, $n_- = \sum_i m_i$, and $\sigma_{(l)}$ is
the dimension of the representation $(l)$ of the permutation group,
which in turn can be computed explicitly as
\begin{equation}
{1\over n_+!}\,\sigma_{(l_1,...,l_s)} = {\prod_{1 \le j \le k \le s} 
(l_j-l_k+k-j)!\over\prod_{i=1}^s(l_i+s-i)!} \,;
\label{sigma-def}
\end{equation}
$d_{(l,m)}$ can be parameterized by
\begin{equation}
d_{(l,m)} = {1\over n_+!}\,{1\over n_-!}\,\sigma_{(l)}\sigma_{(m)}
\,C_{(l,m)},
\end{equation}
where $C_{(l,m)}$ can be expressed as a finite product:
\begin{eqnarray}
C_{(l,m)} &=& \prod_{i=1}^s {(N-t-i+l_i)!\over(N-t-i)!}
\prod_{j=1}^t {(N-s-j+m_j)!\over(N-s-j)!} 
\nonumber \\ &\times&\,
    \prod_{i=1}^s \prod_{j=1}^t {(N+1-i-j+l_i+m_j)!\over(N+1-i-j)!} 
\, ,
\end{eqnarray}
allowing for a conceptually simple $1/N$ expansion.
These results are complemented with the result 
\begin{equation}
F(\beta) = \beta^2 + O(\beta^{2N+2})
\label{F-beta}
\end{equation}
and with the unavoidable large-$N$ constraint $\beta \le
\casefr{1}{2}$.

The character expansion now proceeds as follows.

We notice that, thanks to Eq.\ (\ref{dz}), only a finite number of
nontrivial representations contributes to any definite order in the
strong-coupling series expansion in powers of $\beta$, and each
lattice integration variable can appear only once for each link where
a nontrivial representation in chosen.  A systematic treatment leads
to a classification of contributions in terms of paths (surfaces in a
gauge theory) along whose non self-interacting sections a particular
representation is assigned.  Self-intersection points are submitted to
constraints deriving from the orthogonality of representations and
their composition rules.

In the case of chiral models, all relevant assignments can be
generated by considering the class of the lattice random paths
satisfying a non-backtracking condition
\cite{Campostrini-Rossi-Vicari-chiral-1}.

Once all nontrivial configurations are classified and counted, one is
left with the task of computing the corresponding group integrals.
Only integrations at intersection points are nontrivial, since other
integrations follow immediately from the orthogonality
relationships.  Unfortunately, no special computational
simplifications occur in the large-$N$ limit of group integrals.

Apparently, the character expansion is the most efficient way of
computing the strong-coupling expansion of lattice models.  In
particular, very long strong-coupling series have been obtained in the
large-$N$ limit for the free energy, the mass gap, and the two-point
Green's functions of chiral models in two and three dimensions (for
the free energy, 18 orders on the square lattice, 26 orders on the
honeycomb lattice, and 16 orders on the cubic lattice; for the Green's
functions, 15 orders on the square lattice, 20 orders on the honeycomb
lattice, and 14 orders on the cubic lattice).  The analysis of these
series will be discussed in Sect.\ \ref{principal-chiral}.

Before leaving the present subsection, we must make a few comments
concerning the relationship between ${\rm SU}(N)$ and ${\rm U}(N)$
groups.  We already made the observation that when $N\to\infty$ there
is essentially no difference between ${\rm SU}(N)$ and ${\rm U}(N)$
models, at least when considering operators not involving the
determinant.  In order to explore this relationship more carefully, we
may start as usual from the expression of the single-link integral
(\ref{one-link}).

Representations of $Z(A^\dagger A) $ in the ${\rm SU}(N)$ case can be
obtained \cite{Brower-Rossi-Tan-SUN} in terms of the eigenvalues $x_i$
of $A^\dagger A$ and of $\theta$, defined in Eq.\ (\ref{theta-def}).
Introducing the Vandemonde determinant
\begin{equation}
\Delta(\lambda_1,...,\lambda_N) = \prod_{j>i} (\lambda_j-\lambda_i) =
\det\Vert\lambda_j^{i-1}\Vert,
\end{equation}
one obtains
\begin{eqnarray}
Z(A^\dagger A) &=& {1\over N!} 
\Biggl(\prod_{k=1}^{N-1} {k!\over2\pi}\Biggr)
\int\prod_i\d\phi_i\,\delta\!\left(\sum_i\phi_i + N\theta\right)
\nonumber \\ &\times&\,
{|\Delta(\e^{\I\phi_1},...,\e^{\I\phi_N})|^2 \over
 \Delta(2\sqrt{x_1},...,2\sqrt{x_N})\,
 \Delta(\cos\phi_1,...,\cos\phi_N)}
\exp\Biggl[2\sum_k\sqrt{x_k}\cos\phi_k\Biggr],
\nonumber \\
\end{eqnarray}
or alternatively
\begin{eqnarray}
Z(A^\dagger A) &=& \prod_{k=1}^{N-1} {k!\over2\pi}
\int\prod_i\d\phi_i\,\delta\!\left(\sum_i\phi_i + N\theta\right)
\nonumber \\ &\times&\,
{\Delta(\sqrt{x_1}\e^{\I\phi_1},...,\sqrt{x_N}\e^{\I\phi_N}) \over
 \Delta(x_1,...,x_N)} \exp\Biggl[2\sum_k\sqrt{x_k}\cos\phi_k\Biggr].
\end{eqnarray}

The only difference between ${\rm SU}(N)$ and ${\rm U}(N)$ is due to
the presence of the (periodic) delta function
$\delta\left(\sum_i\phi_i + N\theta\right)$, introducing the
dependence on $\theta$ corresponding to the constraint $\det U = 1$.
A formal solution is obtained by expanding in powers of $\e^{\I
N\theta}$:
\begin{eqnarray}
Z(A^\dagger A) &=& \sum_{m=-\infty}^\infty\e^{\I Nm\theta} 
\det\Vert z_i^{j-1} I_{j-1-|m|}(2z_i)\Vert
\Biggl(\prod_{k=1}^{N-1} k!\Biggr)
{1\over\Delta(z_1^2,...,z_N^2)} \,,
\nonumber \\
\label{Z-sum-m}
\end{eqnarray}
where $z_i = \sqrt{x_i}$.  Eq.\ (\ref{Z-sum-m}) in turn leads to the
following representation of the free energy for the ${\rm SU}(N)$
single-link model:
\begin{equation}
F_N(\beta,\theta) = \log \sum_{m=-\infty}^\infty 
A_{m,N}(\beta) \, \e^{\I Nm\theta},
\label{FN-theta}
\end{equation}
where for convenience we have redefined the coupling:
$\beta\to\beta\e^{\I\theta}$.  Eq.\ (\ref{FN-theta}) is useful for a
large-$N$ mean-field study \cite{Guha-Lee-chiral}, but it is certainly
inconvenient at small $N$, where more specific integration techniques
may be applied.

We mention that a large-$N$ analysis of Eq.\ (\ref{FN-theta}) for
$\theta=0$ leads to
\begin{eqnarray}
F_N(\beta,0) &=& N^2\beta^2 + 2 J_N(2N\beta) - 
2 J_{N-1}(2N\beta)\,J_{N+1}(2N\beta) 
\nonumber \\ &-&\,
\sum_{k=1}^\infty k J^2_{N+k}(2N\beta) + O(\beta^{3N}).
\end{eqnarray}

It is also possible to establish a relationship between ${\rm SU}(N)$
and ${\rm U}(N)$ groups at the level of character coefficients.
Thanks to the basic relationships
\begin{equation}
\chi_{\lambda_1+s,...,\lambda_N+s}(U) = 
(\det U)^s \chi_{\lambda_1,...,\lambda_N}(U),
\end{equation}
holding in ${\rm U}(N)$, one may impose the condition $\det U = 1$ in
the integral representation of the character coefficients and obtain
\begin{equation}
z_{(r)} = {\sum_{s=-\infty}^\infty \tilde z(r,s) \over
           \sum_{s=-\infty}^\infty \tilde z(0,s)},
\end{equation}
where, by definition, for ${\rm U}(N)$ groups
\begin{equation}
\tilde z(0,s) = \left<\det U^s\right>, \qquad
d_{(r)}\,\tilde z(r,s) = 
\left<\det U^s \chi_{(r)}(U)\right>.
\end{equation}
These relationships are the starting point for a systematic
implementation of the corrections due to the ${\rm SU}(N)$ condition
in the $1/N$ expansion of ${\rm U}(N)$ models
\cite{Green-Samuel-chiral,Rossi-Vicari-chiral2}.  A peculiarity of the
${\rm SU}(N)$ condition can be observed in the finite-$N$ behavior of
the eigenvalue density function $\rho(\phi,N)$, which shows a
non-monotonic dependence on $\phi$, characterized by the presence of
$N$ peaks.  This is already apparent in the $\beta\to0$ limit of the
single-link integral, where \cite{Campostrini-Rossi-Vicari-chiral-3}
\begin{eqnarray}
\rho_{{\rm U}(N)}(\phi) \goto_{\beta\to0} {1\over2\pi}\,, &\qquad&
\rho_{{\rm SU}(N)}(\phi) \goto_{\beta\to0} 
{1\over2\pi}\left(1 + (-1)^{N+1}\,{2\over N}\cos N\phi\right).
\nonumber \\
\end{eqnarray}

\section{Chiral chain models and gauge theories on polyhedra}
\label{chiral-chains}

\subsection{Introduction}
\label{sec4intr}

The use of the steepest-descent techniques allows to extend the number
of the unitary-matrix models solved in the large-$N$ limit to some few
unitary-matrix systems.  The interest for few-matrix models may arise
for various reasons. Their large-$N$ solutions may represent
non-trivial benchmarks for new methods meant to investigate the
large-$N$ limit of more complex matrix models, such as QCD.  Every
matrix system may have a r\^ole in the context of two-dimensional
quantum gravity; indeed, via the double scaling limit, its critical
behavior is connected to two-dimensional models of matter coupled to
gravity.  Furthermore, every unitary-matrix model can be reinterpreted
as the generating functional of a class of integrals over unitary
groups, whose knowledge would be very useful for the strong-coupling
expansion of many interesting models.

This section is dedicated to a class of finite-lattice chiral
models termed chain models and defined by the partition function
\begin{equation}
Z_L=\int \prod_{i=1}^L\d U_i \exp\left[
N\beta \sum_{i=1}^L \Tr  \left(U_iU^\dagger_{i+1}
+U^\dagger_iU_{i+1}\right)\right],
\label{Zf}
\end{equation}
where periodic boundary conditions are imposed: $U_{L+1}=U_1$.

Chiral chain models have interesting connections with gauge models.
Fixing the gauge $A_0 = 0$, YM$_2$ on a $K\times L$ lattice (with
free boundary conditions in the direction of size $K$) becomes
equivalent to $K$ decoupled chiral chains of length $L$.

Chiral chains with periodic boundary conditions enjoy another
interesting equivalence with lattice gauge theories defined on the
surface of polyhedra, where a link variable is assigned to each edge
and a plaquette to each face.  By choosing an appropriate gauge,
lattice gauge theories on regular polyhedra like tetrahedron, cube,
octahedron, etc., are equivalent respectively to periodic chiral
chains with $L=4,6,8$, etc.\ \cite{Brower-Rossi-Tan-chains}.

The thermodynamic properties of chiral chains can be derived by
evaluating their partition functions.  Free-energy density, internal
energy, and specific heat are given respectively by
\begin{equation}
F_L={1\over LN^2} \log Z_L,
\label{FL}
\end{equation}
\begin{equation}
U_L= {1\over 2} {\partial F_L\over \partial\beta},
\label{UL}
\end{equation}
\begin{equation}
C_L=\beta^2 {\partial U_L\over \partial \beta}.
\label{CL}
\end{equation}

When $L\rightarrow\infty$, $Z_L$ can be reduced to the partition
function of the Gross-Witten single-link model, and therefore shares
the same thermodynamic properties.  In particular, the free energy
density at $N=\infty$ is piecewise analytic with a third-order
transition at $\beta_c=\casefr{1}{2}$ between the strong-coupling and
weak-coupling domains.  Furthermore, the behavior of $C_\infty$ around
$\beta_c$ can be characterized by a specific heat critical exponent
$\alpha=-1$.  It is easy to see that the $L=2$ chiral chain is also
equivalent to the Gross-Witten model, but with $\beta$ replaced by
$2\beta$; therefore $\beta_c=\casefr{1}{4}$ and the critical
properties are the same, e.g., $\alpha=-1$.

\subsection{Saddle-point equation for chiral $L$-chains} 
\label{exactres}

The strategy used in Refs.\ 
\cite{Brower-Rossi-Tan-chains,Brower-Rossi-Tan-qcd} to compute the
$N=\infty$ solutions for chiral chains with $L\leq 4$ begins with
group integrations in the partition function (\ref{Zf}), with the help
of the single-link integral, for all $U_i$ except two.  This leads to
a representation for $Z_L$ in the form
\begin{equation}
Z_L = \int\d U \, \d V  
\exp \left[N^2 S_{\rm eff}^{(L)}(UV^\dagger)\right]
\label{str1}
\end{equation}
suitable for a large-$N$ steepest-descent analysis.  Since the
integral depends only on the combination $UV^\dagger$, changing
variable to $\theta_j$, $\e^{\I\theta_j}$ being the eigenvalues of
$UV^\dagger$, leads to
\begin{equation}
Z_L \sim \int\prod_i \d\theta_i |\Delta(\theta_1,...,\theta_N)|^2
\exp \left[N^2 S_{\rm eff}^{(L)}(\theta_k)\right]
\label{str2}
\end{equation}
where $-\pi\leq \theta_j \leq \pi$, $\Delta(\theta_1,...,\theta_N) =
\det \Vert\Delta_{jk}\Vert$, $\Delta_{jk}=\e^{\I j\theta_k}$.  In the
large-$N$ limit, $Z_L$ is determined by its stationary configuration,
and the distribution of $\theta_j$ is specified by a density function
$\rho_L(\theta)$, which is the solution of the equation
\begin{equation}
\princint\d\phi\,\rho_L(\phi)
\cot{\theta-\phi\over 2}
+ {\delta\over\delta\theta} S_{\rm eff}^{(L)}(\theta,\rho_L)=0, 
\label{eqrho}
\end{equation}
with the normalization condition
\begin{equation}
\int^{\pi}_{-\pi} \rho_L(\theta)\,\d\theta = 1.
\label{normco}
\end{equation}

For $L=2$, $Z_2$ is already in the desired form with
\begin{equation}
S_{\rm eff}^{(2)} =
2\beta {1\over N} \Tr \left(U_1U_2^\dagger + U_1^\dagger U_2\right),
\label{l2eq1}
\end{equation}
and the large-$N$ eigenvalue density $\rho_2(\theta)$ of the matrix
$U_1U_2^\dagger$ satisfies the Gross-Witten equation
\begin{equation}
\princint\d\phi\,\rho_2(\phi) 
\cot{\theta-\phi\over 2}
-4\beta\sin \theta = 0,
\label{l2eq}
\end{equation}
which differs from that of the infinite-chain model only in replacing
$\beta$ by $2\beta$.

\subsection{The large-$N$ limit of the three-link chiral chain} 
\label{sol3}

In the $L=3$ chain model, setting $U=U_1$ and $V=U_2$, 
$S_{\rm eff}^{(3)}$ is given by
\begin{equation}
\exp\left[ N^2 S_{\rm eff}^{(3)}\right] =
\exp\left[2N\beta \Re \Tr UV^\dagger\right] 
\int\d U_3 \exp\left[2N\beta \Re \Tr A U_3^\dagger\right],
\label{l3eq}
\end{equation}
where $A=U+V$.  Recognizing in the r.h.s.\ of (\ref{l3eq}) a
single-link integral, one can deduce that the large-$N$ limit of the
spectral density $\rho_3(\theta)$ of the matrix $UV^\dagger$ satisfies
the equation
\begin{eqnarray}
2\beta\left(\sin\theta+\sin\half\theta\right)
- \princint\d\phi\,\rho_3(\phi)
\left[\cot{\theta-\phi\over 2} + {1\over 2}
{\sin\half\theta\over \cos\half\theta + \cos\half\phi}
\right] = 0,  \nonumber \\
\label{l3eq2}
\end{eqnarray}
with the normalization condition $\int\rho_3(\theta)\,\d\theta=1$.  In
order to find a solution for the above equation, one must distinguish
between strong-coupling and weak-coupling regions.

In the weak-coupling region the solution of Eq.\ (\ref{l3eq2}) is
\begin{equation}
\rho_3(\theta) = {\beta\over \pi} \cos {\theta\over 4}
\left[2\cos {\theta\over 2} + \sqrt{1 - {1\over 3\beta}}\,\right] 
\left[2\cos {\theta\over 2} - 2\sqrt{1 -{1\over 3\beta}}\,\right]^{1/2}
\label{wesol3}
\end{equation}
for 
\begin{equation}
|\theta| \leq \theta_c = 2\arccos \sqrt{1-{1\over 3\beta}}
\label{thetac}
\end{equation}
and $\rho_3(\theta)=0$ for $\theta_c\leq|\theta|\leq\pi$.  This
solution is valid for $\beta\geq\beta_c=\casefr{1}{3}$, indicating
that a critical point exists at $\beta_c=\casefr{1}{3}$.  Similarly
one can calculate $\rho_3(\theta)$ in the strong-coupling domain
$\beta\leq\beta_c$
\cite{Brower-Rossi-Tan-chains,Brower-Rossi-Tan-qcd,Friedan} finding:
\begin{eqnarray}
\rho_3(\theta) &=& {\beta\over 2\pi} 
\left(y(\theta)+1 - {\sqrt{c} + \sqrt{4+c}\over 2}\right)
\nonumber \\ &\times&
\left[\left(y(\theta)+\sqrt{c}\right)
\left(y(\theta) + \sqrt{4+c}\right)\right]^{1/2},
\label{scsol3}
\end{eqnarray}
where
\begin{equation}
y(\theta) = \sqrt{4\cos^2{\theta\over 2} + c},
\end{equation}
and the parameter $c$ is related to $\beta$ by the equation
\begin{equation}
1 + \sqrt{c} + \casefr{1}{2} c + 
\left(1-\casefr{1}{2}\sqrt{c}\right) \sqrt{4 + c}  
= {1\over\beta} \, .
\label{constr3}
\end{equation}
At $\beta=\beta_c$, $c=0$ and therefore
\begin{equation}
\rho_3(\theta)_{\rm crit}=
{1\over 3\pi} \left(2 \cos{\theta\over 2}\right)^{3/2}
\cos{\theta\over 4},
\label{rho3cr}
\end{equation}
in agreement with the critical limit of the weak-coupling solution
(\ref{wesol3}).

Since $\rho_3(\pi) > 0$ for $\beta < \beta_c$ and $\rho_3(\pi)=0$ for
$\beta\geq \beta_c$, the critical point $\beta_c$ can be also seen as
the compactification point for the spectral density $\rho_3(\theta)$,
similarly to what is observed in the Gross-Witten model.

\subsection{The large-$N$ limit of the four-link chiral chain} 
\label{sol4}

For $L=4$, setting $U=U_1$ and $V=U_3$, $S_{\rm eff}^{(4)}$
is given by
\begin{eqnarray}
\exp\left(N^2 S_{\rm eff}^{(4)}\right) &=&
\int\d U_2\exp\left(2N\beta \Re \Tr A U_2^\dagger\right) \nonumber \\
&\times&
\int\d U_4\exp\left(2N\beta \Re \Tr A U_4^\dagger\right),
\label{l4eq}
\end{eqnarray}
where again $A=U+V$.  The large-$N$ limit of the spectral density
$\rho_4(\theta)$ of the matrix $UV^\dagger$ must be solution of the
equation
\begin{equation}
4\beta\sin\casefr{1}{2}\theta
-\princint\d\phi \rho_4(\phi)
\left[\cot{\theta-\phi\over 2} + 
{\sin\casefr{1}{2}\theta \over 
 \cos\casefr{1}{2}\theta + \cos\casefr{1}{2}\phi}
\right] = 0,
\label{l4eq2}
\end{equation}
satisfying the normalization condition $\int \rho_4(\theta)\d\theta=1$.

In order to solve Eq.\ (\ref{l4eq2}) one must again separate weak- and
strong-coupling domains.  In the weak-coupling region the solution is
\begin{equation}
\renewcommand\arraystretch{1.3}
\begin{array}{l@{\qquad}l}
\displaystyle \rho_4(\theta) = {2\beta\over\pi} 
\sqrt{\sin^2 {\theta_c\over 2} - \sin^2 {\theta\over 2}} &
{\rm for}\ 0\leq \theta\leq\theta_c\leq\pi, \\
\rho_4(\theta) = 0 &
{\rm for}\ \theta_c\leq\theta\leq\pi,
\end{array}
\label{rho4w}
\end{equation}
with $\theta_c$ implicitly determined by the normalization condition
$\int_{-\theta_c}^{\theta_c}\rho_4(\theta)\d\theta=1$.  The solution
(\ref{rho4w}) is valid for $\beta\geq\beta_c=\casefr{1}{8}\pi$, since the
normalization condition can be satisfied only in this region.
$\casefr{1}{8}\pi$ is then a point of non-analyticity representing the
critical point for the transition from the weak to the strong-coupling
domain.

In the strong-coupling domain $\beta < \beta_c=\casefr{1}{8}\pi$ one
finds
\begin{equation}
\rho_4(\theta)={\beta\over 2} 
\sqrt{\lambda - \sin^2 {\theta\over 2}}
\label{rho4s}
\end{equation}
where $\lambda$ is determined by the normalization condition
$\int^\pi_{-\pi}\rho_4(\theta)\d\theta=1$.  The strong- and
weak-coupling expressions of $\rho_4(\theta)$ coincide at $\beta_c$:
\begin{equation}
\rho_4(\theta)_{\rm crit}=
{\beta\over 2} \sqrt{1 - \sin^2 {\theta\over 2}} \, .
\label{rho4cr}
\end{equation}

Notice that again the critical point $\beta_c=\casefr{1}{8}\pi$
represents the compactification point of the spectral density
$\rho_4(\theta)$; indeed $\rho_4(\pi)> 0$ for $\beta< \beta_c$, and
$\rho_4(\pi)=0$ for $\beta\geq \beta_c$.

\subsection{Critical properties of chiral chain models with $L\leq 4$}
\label{crit}

In the following we derive 
the $N=\infty$ critical behavior of the specific heat in the
models with $L=3,4$, using the exact results of
Subs.\ \ref{sol3} and \ref{sol4}.

From the spectral density $\rho_3(\theta)$, the internal energy can be
easily derived by $U_3=\int\d\theta\,\rho_3(\theta) \cos \theta$.  One
finds that $U_3$ is continuous at $\beta_c$.  In the weak-coupling
region $\beta\geq \beta_c=\casefr{1}{3}$,
\begin{eqnarray}
U_3&=& \beta + {1\over 2} - {1\over 8\beta} - 
\beta\left(1 - {1\over 3\beta}\right)^{\!\!3/2},\nonumber \\ 
C_3&=& \beta^2 + {1\over 8} - \beta^2\left(1 + {1\over 6\beta}\right)
\sqrt{1 - {1\over 3\beta}} \, .
\label{ec3w}
\end{eqnarray}
Close to criticality, i.e., for $0\leq \beta/\beta_c-1 \ll 1$,
\begin{equation}
C_3={17\over 72} 
- {1\over 2\sqrt{3}}\left(\beta-\beta_c\right)^{\!\!1/2}
+ O(\beta-\beta_c).
\label{c3cr}
\end{equation}
In the strong-coupling region, one finds 
\begin{equation}
C_3={17\over 72} 
- {1\over 2\sqrt{3}}\left(\beta_c-\beta\right)^{\!\!1/2}
+ O(\beta_c-\beta).
\label{c3crs}
\end{equation}
for $0\leq 1-\beta/\beta_c\ll 1$.  Then the weak- and strong-coupling
expressions of $C_3$ show that the critical point
$\beta_c=\casefr{1}{3}$ is of the third order, and the critical
exponent associated with the specific heat is $\alpha=-\casefr{1}{2}$.

In the $L=4$ case, recalling that $\rho_4(\theta)$ is the spectral
distribution of $U_1 U_3^\dagger$, one writes
\begin{eqnarray}
F_4 &=& {1\over4}\left[
8\beta \int\d\theta\,\rho_4(\theta) \cos{\theta\over 2}
- \int\d\theta\,\d\phi\,\rho_4(\theta)\,\rho_4(\phi)
\log \left(\cos {\theta\over 2}  +\cos {\phi\over 2}\right)\right.
\nonumber \\ &&\quad - \,
\left. {3\over 2} - \log 2\beta 
+\princint\d\theta\,\d\phi\,\rho_4(\theta)\,\rho_4(\phi)
\log \sin^2 {\theta-\phi\over 2}\right] \,.
\label{freen}
\end{eqnarray}
Observing that, since $\rho_4(\theta)$ is a solution of the
variational equation $\delta F_4 / \delta \rho_4 = 0$, the following
relation holds
\begin{equation} 
{\d F_4\over\d\beta} = {\partial F_4\over \partial \beta},
\label{dergfreen}
\end{equation}
one can easily find that
\begin{equation}
U_4 = - {1\over 8\beta} + 
\int\d\theta\,\rho_4(\theta)\cos{\theta\over 2} \, .
\label{en}
\end{equation}

In this case, the study of the critical behavior around
$\beta_c=\casefr{1}{8}\pi$ is slightly subtler, since it requires the
expansion of elliptic integrals $F(k)$ and $E(k)$ around $k=1$.
Approaching criticality from the weak-coupling region, i.e., when
$\beta\rightarrow\beta_c^+$, one obtains
\begin{equation}
C_4={\pi^2\over 32}+{1\over 8} -{\pi^2\over 16\log (4/\delta_w)}
+ O(\delta^2_w) ,
\label{C4w}
\end{equation}
where $\delta^2_w\sim \beta-\beta_c$, apart from logarithms.
For $\beta\rightarrow\beta_c^-$ 
\begin{equation}
C_4={\pi^2\over 32}+{1\over 8} -{\pi^2\over 16\log (4/\delta_s)}
+ O(\delta_s^2) ,
\label{C4s}
\end{equation}
where $\delta_s^2\sim \beta_c-\beta$, apart from logarithms.  A
comparison of Eqs.\ (\ref{C4w}) and (\ref{C4s}) leads to the conclusion
that the phase transition is again of the third order, with a specific
heat critical exponent $\alpha=0^-$.

In conclusion we have seen that chain models with $L=2,3,4,\infty$
have a third-order phase transition at increasing values of the
critical coupling, $\beta_c={1\over4}$, ${1\over3}$,
$\casefr{1}{8}\pi$, ${1\over2}$ respectively, with specific heat
critical exponents $\alpha=-1$, $-\casefr{1}{2}$, $0^-$, $-1$
respectively.  It is worth noticing that $\alpha$ increases when $L$
goes from 2 to 4, reaching the limit of a third order critical
behavior, but in the large-$L$ limit it returns to $\alpha=-1$.

The critical exponent $\nu$, describing the double-scaling behavior
for $N\rightarrow\infty$ and $\beta\rightarrow\beta_c$, can then be
determined by the two-dimensional hyperscaling relationship
$2\nu=2-\alpha$.  This relation has been proved to hold for the
Gross-Witten problem, and therefore for the $L=2$ and $L=\infty$ chain
models, where it is related to the equivalence of the corresponding
double scaling limit with the continuum limit of a two-dimensional
gravity model with central charge $c=-2$.  It is then expected to hold
in general for all values of $L$.  At $L=4$, the value $\nu=1$ has
been numerically verified, within a few per cent of uncertainty, by
studying the scaling of the specific heat peak position at finite $N$.
Notice that the exponents $\alpha=0^-$, $\nu=1$ found for $L=4$
correspond to a central charge $c=1$.

\subsection{Strong-coupling expansion of chiral chain models}
\label{SC}

Strong-coupling series of the free energy density of chiral chain
models can be generated by means of the character expansion, which
leads to the result
\begin{equation}
F_L(\beta) =  F(\beta) + \widetilde{F}_L(\beta),
\label{be1}
\end{equation}
where $F(\beta)$ is the free energy of the single unitary-matrix model,
\begin{equation}
\widetilde{F}_L = {1\over LN^2}\log \sum_{(r)}d_{(r)}^2 z_{(r)}^L,
\label{be1b}
\end{equation}
$\sum_{(r)}$ denotes the sum over all irreducible representations of
${\rm U}(N)$, and $d_{(r)}$ and $z_{(r)}(\beta)$ are the corresponding
dimensions and character coefficients.  The calculation of the
strong-coupling series of $F_L(\beta)$ is considerably simplified in
the large-$N$ limit, due to the relationships (\ref{F-beta}) and 
\begin{equation}
z_{(r)}(\beta) =  
\bar{z}_{(r)} \beta^n + O\left(\beta^{2N}\right),
\label{be4}
\end{equation}
where $\bar{z}_{(r)}$ is independent of $\beta$ and $n$ is the order
of the representation $(r)$.  Explicit expressions for $d_{(r)}$ and
$\bar{z}_{(r)}$ were reported in Subs.\ \ref{character-expansion}.
The large-$N$ strong-coupling expansion of $\widetilde{F}_L(\beta)$ is
actually a series in $\beta^L$, i.e.,
\begin{equation}
\widetilde{F}_L = \sum_n c(n,L)\beta^{nL}.
\label{be5}
\end{equation}
It is important to recall that the large-$N$ character coefficients
have jumps and singularities at $\beta={1\over 2}$
\cite{Green-Samuel-chiral}, and therefore the relevant region for a
strong-coupling character expansion is $\beta<{1\over 2}$.

Another interesting aspect of the large-$N$ limit of chain models,
studied by Green and Samuel using the strong-coupling character
expansion \cite{Green-Samuel-un2}, concerns the determinant channel,
which should provide an order parameter for the phase transition.  The
quantity
\begin{equation}
\sigma = -{1\over N} \log \langle \det U_i U_{i+1}^\dagger \rangle
\label{detch}
\end{equation}
is non-zero in the strong-coupling domain and zero in weak coupling at
$N=\infty$.  $\beta_c$ may then be evaluated by determining where the
strong-coupling evaluation of the order parameter $\sigma$ vanishes.
Like the free-energy, $\sigma$ is calculable via a character
expansion.  Indeed
\begin{equation}
\langle \det U_i U_{i+1}^\dagger \rangle =
{ \sum_{(r)} d_{(r)} z_{(r)}^{L-1} d_{(r,-1)} z_{(r,-1)} \over 
\sum_{(r)} d_{(r)}^2 z_{(r)}^L}
\label{scdet}
\end{equation}
Green and Samuel evaluated a few orders of the above character
expansion, obtaining estimates of $\beta_c$ from the vanishing point
of $\sigma$. Such estimates compare well with the exact results for
$L=3,4$. In the cases where $\beta_c$ is unknown, they found
$\beta_c\simeq 0.44$ for $L=5$, $\beta_c\simeq 0.47$ for $L=6$, etc.,
with $\beta_c$ monotonically approaching the value $\casefr{1}{2}$
with increasing $L$.

In order to study the critical behavior of chain models for $L\geq 5$,
one can also analyze the corresponding strong-coupling series of the
free energy (\ref{be1})
\cite{Brower-Campostrini-Orginos-Rossi-Tan-Vicari}.  An integral
approximant analysis of the strong-coupling series of the specific
heat led to the estimates $\beta_c\simeq 0.438$ for $L=5$ and
$\beta_c\simeq 0.474$ for $L=6$, with small negative $\alpha$, which
could mimic an exponent $\alpha=0^-$. For $L\geq 7$ a such
strong-coupling analysis would lead to $\beta_c$ larger than
$\casefr{1}{2}$, that is out of the region where a strong-coupling
analysis can be predictive.  Therefore something else must occur
earlier, breaking the validity of the strong-coupling expansion.  An
example of this phenomenon is found in the Gross-Witten single-link
model (recovered when $L\rightarrow\infty$), where the strong-coupling
expansion of the $N=\infty$ free energy is just $F(\beta) = \beta^2$,
an analytical function without any singularity; therefore, in this
model, $\beta_c={1\over 2}$ cannot be determined from a
strong-coupling analysis of the free energy.

From such analysis one may hint at the following possible scenario: as
for $L\leq4$, for $L=5,6$, that is when the estimate of $\beta_c$
coming from the above strong-coupling analysis is smaller than
${1\over 2}$ and therefore acceptable.  The term
$\widetilde{F}(\beta)$ in Eq.\ (\ref{be1}) should be the one relevant
for the critical properties, determining the critical points and
giving $\alpha\neq -1$ (maybe $\alpha=0^-$ as in the $L=4$ case).  For
$L\geq 7$ the critical point need not be a singular point of the free
energy in strong or weak coupling, but just the point where
weak-coupling and strong-coupling curves meet each other.  This would
cause a softer phase transition with $\alpha=-1$, as for the
Gross-Witten single-link problem.  We expect $\beta_c<{1\over 2}$ also
for $L\geq 7$.  This scenario is consistent with the results of the
analysis of the character expansion of $\sigma$, defined in Eq.\ 
(\ref{detch}).

\section{Simplicial chiral models}
\label{simplicial-chiral}

\subsection{Definition of the models}
\label{simplicial-def}

Another interesting class of finite-lattice chiral models is obtained
by considering the possibility that each of a finite number of unitary
matrices may interact in a fully symmetric way with all other
matrices, while preserving global chiral invariance; the resulting
systems can be described as chiral models on $(d-1)$-dimensional
simplexes, and thus termed ``simplicial chiral models''
\cite{Brower-Campostrini-Orginos-Rossi-Tan-Vicari,Rossi-Tan}.

The partition function for such a system is:
\begin{equation}
Z_d = \int\prod_{i=1}^d \d U_i \,
\exp\Biggl[N\beta\sum_{i=1}^d\sum_{j=i+1}^d 
\Tr\left(U_i U_j^\dagger + U_j U_i^\dagger\right)\Biggr].
\label{Z-simplicial}
\end{equation}
Eq.~(\ref{Z-simplicial}) encompasses as special cases a number of
models that we have already introduced and solved; in particular, the
chiral chains with $L\le3$ correspond to the simplicial chiral models
with $d\le3$.

One of the most attractive features of these models is their
relationship with higher-dimensional systems, with which they share
the possibility of high coordination numbers.  This relationship
becomes exact in the large-$d$ limit, where mean-field results are
exact.

In the large-$N$ limit and for arbitrary $d$ a saddle-point equation
can be derived, whose solution allows the evaluation of the large-$N$
free energy
\begin{equation}
F_d = {1\over N^2}\,\log Z_d
\end{equation}
and of related thermodynamical quantities.

\subsection{Saddle-point equation for simplicial chiral models}
\label{simplicial-saddle}

The strategy for the determination of the large-$N$ saddle-point
equation is based on the introduction of a single auxiliary variable
$A$ (a complex matrix), allowing for the decoupling of the unitary
matrix interaction:
\begin{equation}
Z_d = {\widetilde Z_d\over \widetilde Z_0} \,,
\end{equation}
where
\begin{eqnarray}
\widetilde Z_d = \int\prod_{i=1}^d \d U_i\,\d A \,
\exp\Biggl[- N\beta && \Tr AA^\dagger + N\beta\Tr A\sum_i U_i^\dagger
\nonumber \\ + N\beta && \Tr A^\dagger\sum_i U_i - N^2\beta d\Biggr].
\end{eqnarray}

We are now back to the single-link problem and, since we have solved
it in Sect.\ \ref{single-link} in terms of the function $W$, whose
large-$N$ limit is expressed by Eq.~(\ref{SD-W}), we obtain
\begin{equation}
\widetilde Z_d = \int \d A \,
\exp\left[-N\beta \Tr A A^\dagger + N d W(\beta^2 A A^\dagger)
    - N^2 \beta d\right].
\end{equation}
It is now convenient to express the result in terms of the eigenvalues
$x_i$ of the Hermitian semipositive-definite matrix $4 \beta A
A^\dagger$, obtaining
\begin{equation}
\widetilde Z_d = \int \d\mu(x_i) \, 
\exp\Biggl[-{N\over4\beta} \sum_i x_i + N d W\left(x_i\over4\right)
    - N^2 \beta d\Biggr].
\end{equation}
The angular integration can be performed, leading to
\begin{equation}
\d\mu(x_i) = \prod_i \d x_i \prod_{i>j} (x_i-x_j)^2.
\end{equation}
The saddle-point equation is therefore 
\begin{equation}
{\sqrt{r+x_i}\over2\beta} - d =
{1\over N}\sum_{i\ne j}
{(4-d)\sqrt{r+x_i} + d \sqrt{r+x_j} \over x_i-x_j} \, ,
\label{simpl-SP}
\end{equation}
subject to the constraint (needed to define $r$)
\begin{equation}
\left\{
\renewcommand\arraystretch{1.3}
\begin{array}{l@{\quad}l}
\displaystyle {1\over N}\sum_i {1\over\sqrt{r+x_i}} = 1 &
\hbox{(strong coupling)}; \\
r = 0 & \hbox{(weak coupling)}.
\end{array}
\right.
\end{equation}

The energy
\begin{equation}
U_d = {1\over2}\,{\partial F_d\over\partial\beta}
\end{equation}
is easily expressed in terms of the eigenvalues:
\begin{equation}
d(d-1)U_d = {1\over4\beta^2} \sum_i x_i - d - {1\over\beta} \,.
\label{simpl-Ud}
\end{equation}

In the large-$N$ limit, after a change of variables to
$z_i=\sqrt{r+x_i}$, we introduce as usual an eigenvalue density
function $\rho(z)$, and turn Eq.~(\ref{simpl-SP}) into the integral
equation 
\begin{equation}
{z\over2\beta} - d = \princint_a^b \d z' \, \rho(z')
\left[{2\over z-z'} - {d-2\over z+z'}\right],
\label{simpl-IE}
\end{equation}
subject to the constraints
\begin{equation}
\int_a^b \rho(z') \,\d z' = 1
\end{equation}
and
\begin{equation}
\int_a^b \rho(z') \, {\d z'\over z'} \le 1,
\end{equation}
with equality holding in strong coupling, where $a=\sqrt{r}$.
The easiest way of evaluating the free energy $F_d$ is the integration
of the large-$N$ version of Eq.\ (\ref{simpl-Ud}) with respect to
$\beta$.

Very simple solutions are obtained for a few special values of $d$.
When $d=0$, the problem reduces to a Gaussian integration, and one
easily finds that Eq.\ (\ref{simpl-IE}) is solved by
\begin{equation}
\rho(z) = {z\over4\pi\beta}\,
{\sqrt{16\beta - (z^2-a^2)}\over\sqrt{z^2-a^2}}
\end{equation}
and $\widetilde Z_0 = \exp(N^2\log\beta)$, independent of $a$ as
expected.  

When $d=2$ we obtain
\begin{eqnarray}
\rho_w(z) &=& {1\over4\pi\beta}\sqrt{8\beta-(z-4\beta)^2}, \qquad
\beta\ge\casefr{1}{2}, \\
\rho_s(z) &=& {1\over4\pi\beta} \, z\sqrt{1+6\beta-z\over z-(1-2\beta)} \,,
\qquad r(\beta) = (1-2\beta)^2,
\qquad \beta\le\casefr{1}{2}, 
\end{eqnarray}
and these results are consistent with the reinterpretation of the
model as a Gross-Witten one-plaquette system.  Notice however that the
matrix whose eigenvalue distribution has been evaluated is not the
original unitary matrix, and corresponds to a different choice of
physical degrees of freedom.  This is the reason why, while knowing
the solution for the free energy of the $d=1$ system (trivial,
non-interacting) and of the $d=3$ system (three-link chiral chain), we
cannot find easily explicit analytic forms for the corresponding
eigenvalue densities.

The saddle-point equation (\ref{simpl-IE}) has been the subject of
much study in recent times, because it is related to many different
physical problems in the context of double-scaling limit
investigations.  In particular, in the range of values $0\le d\le4$,
the same equation describes the behavior of ${\rm O}(n)$ spin models
on random surfaces in the range $-2\le n\le2$, with the very simple
mapping $n=d-2$ \cite{Gaudin-Kostov}. In this range, the equation has
been solved analytically in Refs.\ \cite{Eynard-Kristjansen-1} and
especially \cite{Eynard-Kristjansen-2} in terms of $\theta$-functions.

\subsection{The large-$N$ $d=4$ simplicial chiral model}
\label{simpl-d=4}

The chiral model on a tetrahedron is the first example within the
family of simplicial chiral models which turns out to be really
different from all the systems discussed in the previous sections.
Explicit solutions were found for both the weak and the strong
coupling phases, and they are best expressed in terms of a rescaled
variable
\begin{equation}
\zeta = \sqrt{1 - {z^2\over b^2}}
\end{equation}
and of a dynamically determined parameter
\begin{equation}
k = \sqrt{1 - {a^2\over b^2}} \, .
\end{equation}
The resulting expressions, after defining 
$\beta\bar{\rho}(\zeta)\,\d\zeta \equiv \rho(z)\,\d z$, are
\begin{equation}
\bar{\rho}_w(\zeta) = {8\over E(k)^2}
\left[{\sqrt{k^2-\zeta^2}\over\sqrt{1-\zeta^2}}\,K(k) -
       \sqrt{k^2-\zeta^2}\sqrt{1-\zeta^2}\,\Pi(\zeta^2,k)\right]
\end{equation}
and
\begin{eqnarray}
\bar{\rho}_s(\zeta) &=& {8\over[E(k)-(1-k^2)K(k)]^2} \nonumber \\
&\times&
\left[k^2\,{\sqrt{1-\zeta^2}\over\sqrt{k^2-\zeta^2}}\,K(k) -
       \sqrt{k^2-\zeta^2}\sqrt{1-\zeta^2}\,\Pi(\zeta^2,k)\right],
\end{eqnarray}
where $K$, $E$ and $\Pi$ are the standard elliptic integrals, and
$0\le\zeta\le k$.  

The complete solution is obtained by enforcing the normalization
condition, which leads to a relationship between $\beta$ and $k$, best
expressed by the equation
\begin{equation}
{1\over\beta} = \int_0^k \d\xi\,\bar{\rho}(\zeta,k).
\end{equation}
Criticality corresponds to the limit $k\to1$, and it is easy to
recognize that both weak and strong coupling results lead in this
limit to $\beta_c=\casefr{1}{4}$ and
\begin{equation}
\beta\bar{\rho}_c(\zeta) = \zeta\log{1+\zeta\over1-\zeta}\,.
\end{equation}

Many interesting features of this model in the region around
criticality can be studied analytically, and one may recognize that
the critical behavior around $\beta_c=\casefr{1}{4}$ corresponds to a
limiting case of a third-order phase transition with critical exponent
of the specific heat $\alpha=0^-$.  In the double-scaling limit
language this would correspond to a model with central charge $c=1$
and logarithmic deviations from scaling.  The critical behavior of the
specific heat on both sides of criticality is described by
\begin{equation}
C \equiv \beta^2\,{\partial U\over\partial\beta}
\goto_{k'\to0} 
{\pi^2 + 3\over36} - {\pi^2\over12\log(4/k')} 
+ O\left(1\over\log^2k'\right),
\end{equation}
where $k'\equiv\sqrt{1-k^2}$.

\subsection{The large-$d$ limit}
\label{simpl-large-d}

By introducing a function defined by
\begin{equation}
f(z) = \int_a^b {\rho(z')\over z-z'}\,\d z', \qquad
f(z) \goto_{|z|\to\infty} {1\over z}\,,
\label{simpl-f}
\end{equation}
analytic in the complex $z$ plane with the exception of a cut on the
positive real axis in the interval $[a,b]$, we can turn the
saddle-point equation (\ref{simpl-IE}) into the functional equation
\begin{equation}
{z\over2\beta} - d = 2 \Re f(z) + (d-2) f(-z).
\label{simpl-FE}
\end{equation}

This equation can be the starting point of a systematic $1/d$
expansion, on whose details we shall not belabor, especially because
its convergence for small values of $d$ is very slow.
It is however interesting to solve the large-$d$ limit of 
Eq.\ (\ref{simpl-FE}) by the Ansatz
\begin{equation}
\rho(z) = \delta(z-\bar z),
\end{equation}
whose substitution into Eq.\ (\ref{simpl-f}) leads to the solution
\begin{equation}
\renewcommand\arraystretch{1.3}
\begin{array}{l@{\qquad}l}
\displaystyle \bar z = \beta d
\left(1 + \sqrt{1 - {1\over\beta d}}\right), &
\beta d\ge1, \\
\bar z = 1, &
\beta d\le1.
\end{array}
\end{equation}

The large-$d$ limit predicts the location of the critical point
$\beta_c = 1/d$, and shows complete equivalence with the mean-field
solution of infinite-volume principal chiral models on a
$d/2$-dimensional hypercubic lattice.  The large-$d$ prediction for
the nature of criticality is that of a first-order phase transition,
with
\begin{equation}
U = {1\over2} + {1\over2}\sqrt{1 - {1\over\beta d}} - 
{1\over4\beta d} \,, \qquad \beta d \ge 1.
\end{equation}

\subsection{The large-$N$ criticality of simplicial models}
\label{simpl-crit}

The connection with the double-scaling limit problem naturally leads
to the study of the finite-$\beta$ critical behavior.  In the regime
$0\le d\le4$ one is helped by the equivalence with the solved problem
of ${\rm O}(n)$ spin models on a random surface, which allows not only
a determination of the critical value (found to satisfy the
relationship $\beta_c d=1$), but also an evaluation of the eigenvalue
distribution at criticality \cite{Gaudin-Kostov}:
\begin{equation}
\rho_c(z) = {2\over\pi\theta}\,\cos{\pi\theta\over2}\,
{\sinh\theta u\over\cosh u}\,,
\end{equation}
and
\begin{equation}
a_c = 0, \qquad b_c = {2\over\theta}\,\tan{\pi\theta\over2} \,,
\end{equation}
where $\theta$ and $u$ are defined by the parametrizations
\begin{equation}
4\cos^2{\pi\theta\over2} \equiv d = {1\over\beta_c} \,, \qquad
\cosh u \equiv {b_c\over z} \,.
\end{equation}

Unfortunately, the technique that was adopted in order to find the
above solution does not apply to the regime $d>4$, in which case one
cannot choose $a_c=0$.  The saddle-point equation at criticality can
however be solved numerically with very high accuracy, and one finds
that the relationship
\begin{equation}
\beta_c d = 1
\end{equation}
is satisfied for all $d$, thus also matching the large-$d$
predictions.  The combinations $(a_c+b_c)/2$ and $a_cb_c$ admit a
$1/d$ expansion, and the coefficients of the expansion are found
numerically to be integer numbers up to order $d^{-8}$.

An analysis of criticality for $d>4$ shows that its description is
fully consistent with the existence of a first-order phase transition,
with a discontinuity of the internal energy measured by 
$d a_c^2/(4(d-1))$, again matching with the large-$d$ (mean-field)
predictions.

\subsection{The strong-coupling expansion of simplicial models}
\label{simpl-sc}

There is nothing peculiar in performing the strong-coupling expansion
of Eq.\ (\ref{Z-simplicial}).  There is however a substantial
difference with respect to the case of chiral chains discussed in the
previous section: because of the topology of simplexes, the
strong-coupling configurations entering the calculation are no longer
restricted to simple graphs whose vertices are joined by at most one
link, and the full complexity of group integration on arbitrary graphs
is now involved \cite{Campostrini-Rossi-Vicari-chiral-1}.

As a consequence, as far as the simplicial models can be solved by
different techniques, they may also be used as generating functionals
for these more involved group integrals, that enter in a essential way
in all strong-coupling calculations in higher-dimensional standard
chiral models and lattice gauge theories.

\section{Asymptotically free matrix models}
\label{principal-chiral}

\subsection{Two-dimensional principal chiral models}
\label{sec6intr}

Two dimensional ${\rm SU}(N)\times {\rm SU}(N)$ principal chiral
models, defined by the action
\begin{equation}
S={1\over T} \int\d^2x \Tr\partial_\mu U(x) \, \partial_\mu
U^\dagger(x),
\label{caction}
\end{equation}
are the simplest asymptotically free field theories whose large-$N$
limit is a sum over planar diagrams, like four dimensional ${\rm
SU}(N)$ gauge theories.

Using the existence of an infinite number of conservation laws and
Bethe-Ansatz methods, the on-shell solution of the ${\rm SU}(N)\times
{\rm SU}(N)$ chiral models has been proposed in terms of a factorized
$S$-matrix \cite{Abdalla-Abdalla-LimaSantos,Wiegmann}.  The analysis
of the corresponding bound states leads to the mass spectrum
\begin{equation}
M_r=M{\sin(r\pi/N) \over \sin(\pi/N)},
\qquad 1\leq r\leq N-1,
\label{masses}
\end{equation}
where $M_r$ is the mass of the $r$-particle bound state transforming
as totally antisymmetric tensors of rank $r$.  $M\equiv M_1$ is the
mass of the fundamental state determining the Euclidean long-distance
exponential behavior of the two-point Green's function
\begin{equation}
G(x)= {1\over N}\langle {\rm Tr} \,U(0) U(x)^\dagger \rangle.
\label{fgf}
\end{equation} 
The mass-spectrum (\ref{masses}) has been verified numerically at
$N=6$ by Monte Carlo simulations
\cite{Rossi-Vicari-chiral1,Drummond-Horgan}: Monte Carlo data of the
mass ratios $M_2/M$ and $M_3/M$ agree with formula (\ref{masses})
within statistical errors of about one per cent.  Concerning the
large-$N$ limit of these models, it is important to notice that the
$S$-matrix has a convergent expansion in powers of $1/N$, and becomes
trivial, i.e., the $S$-matrix of free particles, in the large-$N$
limit.

By using Bethe-Ansatz techniques, the mass/$\Lambda$-parameter
ratio has also been computed, and the result is
\cite{Balog-Naik-Niedermayer-Weisz}
\begin{equation}
{M\over\Lambda_{\overline {MS}}}=\sqrt{{8\pi\over\e}} 
\, {\sin (\pi/N)\over \pi/N},
\label{mass-lambda}    
\end{equation}
which again enjoys a $1/N$ expansion with a finite radius of
convergence. This exact but non-rigorous result has been substantially
confirmed by Monte Carlo simulations at several values of $N$
\cite{Rossi-Vicari-chiral2,Manna-Guttmann-Hughes}, and its large-$N$
limit also by $N=\infty$ strong-coupling
calculations \cite{Campostrini-Rossi-Vicari-chiral-brief,%
Campostrini-Rossi-Vicari-chiral-2}.

While the on-shell physics of principal chiral models has been
substantially solved, exact results of the off-shell physics are still
missing, even in the large-$N$ limit.  When $N\rightarrow\infty$,
principal chiral models should just reproduce a free-field theory in
disguise.  In other words, a local nonlinear mapping should exist
between the Lagrangian fields $U$ and some Gaussian variables
\cite{Polyakov-book}.  However, the behavior of the two-point Green's
function $G(x)$ of the Lagrangian field shows that such realization of
a free-field theory is nontrivial.  While at small Euclidean momenta,
and therefore at large distance, there is a substantial numerical
evidence for an essentially Gaussian behavior of $G(x)$
\cite{Rossi-Vicari-chiral2}, at short distance renormalization group
considerations lead to the asymptotic behavior
\begin{equation}
G(x) \sim
\left[\log\left({1\over x \Lambda}\right)\right]^{\gamma_1/b_0} ,
\label{n14}
\end{equation}
where $\Lambda$ is a mass scale, and
\begin{equation}
{\gamma_1 \over b_0} = 2\left(1 -{2\over N^2}\right)
\goto_{N\rightarrow\infty} 2.
\end{equation}
$b_0$ and $\gamma_1$ are the first coefficients respectively of the
$\beta$-function and of the anomalous dimension of the fundamental
field.  We recall that a free Gaussian Green's function behaves like
$\log\left(1/x \right)$.  Then at small distance $G(x)$ seems to
describe the propagation of a composite object formed by two
elementary Gaussian excitations, suggesting an interesting
hadronization picture: in the large-$N$ limit, the Lagrangian fields
$U$, playing the r\^ole of non-interacting hadrons, are constituted by
two confined particles, which appear free in the large momentum limit,
due to asymptotic freedom.

Numerical investigations by Monte Carlo simulations of lattice chiral
models in the continuum limit show that the large-$N$ limit is rapidly
approached, which confirms that the $1/N$ expansion, were it
available, would be an effective predictive tool in the analysis of
these models.

\subsection{Principal chiral models on the lattice}
\label{sec6s2}

In the persistent absence of an explicit solution, the large-$N$ limit
of two-dimensional chiral models has been investigated by applying
analytical and numerical methods of lattice field theory, such as
strong-coupling expansion and Monte Carlo simulations.  In the
following we describe the main results achieved by these studies.

A standard lattice version of the continuum action (\ref{caction}) is
obtained by introducing a nearest-neighbor interaction, according to
Eq.\ (\ref{action-spin}):
\begin{equation}
S_L=-2N\beta\sum_{x,\mu} 
{\rm Re}{\rm Tr}\left[ U_x U^\dagger_{x+\mu}\right], 
\qquad \beta={1\over NT}\,.
\label{laction}
\end{equation}
${\rm SU}(N)$ and ${\rm U}(N)$ lattice chiral models, obtained by
constraining respectively $U_x\in {\rm SU}(N)$ and $U_x\in {\rm
U}(N)$, are expected to have the same large-$N$ limit at fixed
$\beta$.  In the continuum limit $\beta\rightarrow\infty$, ${\rm
SU}(N)$ and ${\rm U}(N)$ lattice actions should describe the same
theory even at finite $N$, since the additional ${\rm U}(1)$ degrees
of freedom of ${\rm U}(N)$ models should decouple.  In other words,
the ${\rm U}(N)$ lattice theory represents a regularization of the
${\rm SU}(N)\times {\rm SU}(N)$ chiral field theory when restricting
ourselves to its ${\rm SU}(N)$ degrees of freedom, i.e. when
considering Green's functions of the field
\begin{equation}
\hat{U}_x = {U_x\over(\det U_x)^{1/N}} \, , 
\label{Uhat}
\end{equation}
e.g.,
\begin{equation}
G(x)\equiv {1\over N} 
\langle {\rm Tr} \hat{U}_0 \hat{U}_x^\dagger\rangle, 
\end{equation}
whose large-distance behavior allows to define the fundamental mass
$M$.

At finite $N$, while ${\rm SU}(N)$ lattice models should not have any
singularity at finite $\beta$, ${\rm U}(N)$ lattice models should
undergo a phase transition, driven by the ${\rm U}(1)$ degrees of
freedom corresponding to the determinant of $U(x)$.  The determinant
two-point function
\begin{equation}
G_d(x)\equiv \langle \det [ U^\dagger (x) U(0) ]\rangle^{1/N}
\end{equation}
behaves like $x^{-f(\beta,N)}$ at large $x$ in the weak-coupling
region, with $f(\beta,N)\sim O(1/N)$, but drops off exponentially in
strong-coupling region, where $G_d(x)\sim\e^{-m_d x}$ with
\cite{Green-Samuel-un2}
\begin{equation}
m_d=-\log\beta + {1\over N}\log{N!\over N^N} + O(\beta^2).
\label{scmd}
\end{equation}
This would indicate the existence of a phase transition at a finite
$\beta_d$ in ${\rm U}(N)$ lattice models.  Such a transition, being
driven by ${\rm U}(1)$ degrees of freedom, should be of the
Kosterlitz-Thouless type: the mass propagating in the determinant
channel $m_d$ should vanish at the critical point $\beta_d$ and stay
zero for larger $\beta$.  Hence for $\beta > \beta_d$ this ${\rm
U}(1)$ sector of the theory would decouple from the ${\rm SU}(N)$
degrees of freedom, which alone determine the continuum limit
($\beta\to\infty$) of principal chiral models.

The large-$N$ limit of principal chiral models has been investigated
by Monte Carlo simulations of ${\rm SU}(N)$ and ${\rm U}(N)$ models
for several large values of $N$, studying their approach to the
$N=\infty$
limit \cite{Rossi-Vicari-chiral2,Campostrini-Rossi-Vicari-chiral-3}.

Many large-$N$ strong-coupling calculations have been performed which
allow a direct study of the $N=\infty$ limit.  Within the
nearest-neighbor formulation (\ref{laction}), the large-$N$
strong-coupling expansion of the free energy has been calculated up to
18th order, and that of the fundamental Green's function $G(x)$
(defined in Eq.\ (\ref{fgf})) up to 15th order
\cite{Green-Samuel-un2,Campostrini-Rossi-Vicari-chiral-1}.  Large-$N$
strong-coupling calculations have been performed also on the honeycomb
lattice, within the corresponding nearest-neighbor formulation, which
is expected to belong to the same class of universality with respect
to the critical point $\beta=\infty$. On the honeycomb lattice the
free energy has been computed up to $O\left(\beta^{26}\right)$, and
$G(x)$ up to $O\left(\beta^{20}\right)$
\cite{Campostrini-Rossi-Vicari-chiral-1}.

Monte Carlo simulations show that ${\rm SU}(N)$ and ${\rm U}(N)$
lattice chiral models have a peak in the specific heat
\begin{equation}
C= {1\over N} {\d E\over\d T} 
\end{equation}
which becomes sharper and sharper with increasing $N$, suggesting the
presence of a critical phenomenon for $N=\infty$ at a finite
$\beta_c$.  In ${\rm U}(N)$ models the peak of $C$ is observed in the
region where the determinant degrees of freedom are massive, i.e., for
$\beta < \beta_d$ (this feature characterizes also
two-dimensional ${\rm XY}$ lattice models \cite{Gupta-Baillie}).  An
estimate of the critical coupling $\beta_c$ has been obtained by
extrapolating the position $\beta_{\rm peak}(N)$ of the peak of the
specific heat (at infinite volume) to $N\rightarrow\infty$ using a
finite-$N$ scaling Ansatz \cite{Campostrini-Rossi-Vicari-chiral-3}
\begin{equation}
\beta_{\rm peak}(N) \simeq \beta_c+ cN^{-\epsilon},
\label{FNS}
\end{equation}
mimicking a finite-size scaling relationship.  The above Ansatz arises
from the idea that the parameter $N$ may play a r\^ole quite analogous
to the volume in the ordinary systems close to the criticality.  This
idea was already exploited in the study of one-matrix models
\cite{Damgaard-Heller,Carlson,Brezin-ZinnJustin-RG}, where the double
scaling limit turns out to be very similar to finite-size scaling in a
two-dimensional critical phenomenon.  The finite-$N$ scaling Ansatz
(\ref{FNS}) has been verified in the similar context of the large-$N$
Gross-Witten phase transition, as mentioned in Subs.\  
\ref{double-scaling-single-link}.  Since $\epsilon$ is supposed to be
a critical exponent associated with the $N=\infty$ phase transition,
it should be the same in the ${\rm U}(N)$ and ${\rm SU}(N)$ models.

The available ${\rm U}(N)$ and ${\rm SU}(N)$ Monte Carlo data (at
$N=9,15,21$ for ${\rm U}(N)$ and $N=9,15,21,30$ for ${\rm SU}(N)$) fit
very well the Ansatz (\ref{FNS}), and their extrapolation leads to the
estimates $\beta_c= 0.3057(3)$ and $\epsilon=1.5(1)$.  The
interpretation of the exponent $\epsilon$ in this context is still an
open problem.  It is worth noticing that the value of the correlation
length describing the propagation in the fundamental channel is finite
at the phase transition: $\xi^{(c)}\simeq 2.8$.

The existence of this large-$N$ phase transition is confirmed by an
analysis of the $N=\infty$ 18th-order strong-coupling series of the
free energy
\begin{eqnarray}
F=&&2\beta^2+2\beta^4+4\beta^6+19\beta^8+96\beta^{10}+
604\beta^{12}\nonumber \\ &&+4036\beta^{14}
+ {58471\over 2}\beta^{16}+{663184\over 3}\beta^{18}+
O\left(\beta^{20}\right),
\label{Fseries}
\end{eqnarray}
which shows a second-order critical behavior:
\begin{equation}
C ={1\over 4} \beta^2 {\partial^2 F\over \partial\beta^2}
\sim  |\beta - \beta_c |^{-\alpha},
\label{Ccrit}
\end{equation}
with $\beta_c = 0.3060(4)$ and $\alpha = 0.27(3)$, in agreement with
the extrapolation of Monte Carlo data.  The above estimates of
$\beta_c$ and $\alpha$ are slightly different from those given in
Ref.\ \cite{Campostrini-Rossi-Vicari-chiral-2}; they are obtained by a
more refined analysis based on integral approximant techniques
\cite{Guttmann-Joyce,Hunter-Baker-AI,Fisher-Yang} and by the so-called
critical point renormalization method \cite{Hunter-Baker-CPRM}.

Green and Samuel argued that the large-$N$ phase transition of
principal chiral models on the lattice is nothing but the large-$N$
limit of the determinant phase transition present in ${\rm U}(N)$
lattice models \cite{Green-Samuel-chiral,Green-Samuel-largeN}.
According to this conjecture, $\beta_d$ and $\beta_{\rm peak}$
should both converge to $\beta_c$ in the large-$N$ limit, and the
order of the determinant phase transition would change from the
infinite order of the Kosterlitz-Thouless mechanism to a second order
with divergent specific heat.  The available Monte Carlo data of ${\rm
U}(N)$ lattice models at large $N$ provide only a partial confirmation
of this scenario; one can just get a hint that $\beta_d(N)$ is
also approaching $\beta_c$ with increasing $N$.  The large-$N$ phase
transition of the ${\rm SU}(N)$ models could then be explained by the
fact that the large-$N$ limit of the ${\rm SU}(N)$ theory is the same
as the large-$N$ limit of the ${\rm U}(N)$ theory.

The large-$N$ character expansion of the mass $m_d$ propagating
in the determinant channel has been calculated up to 6th order in the
strong-coupling region, indicating a critical point (determined by the
zero of the $m_d$ series) slightly larger than our determination
of $\beta_c$: $\beta_d(N{=}\infty)\simeq 0.324$
\cite{Green-Samuel-chiral}.  This discrepancy might be explained
either by the shortness of the available character expansion of
$m_d$ or by the fact that such a determination of $\beta_c$
relies on the absence of singular points before the strong-coupling
series of $m_d$ vanishes, and therefore a non-analyticity at
$\beta_c\simeq 0.306$ would invalidate all strong-coupling predictions
for $\beta > \beta_c$.

It is worth mentioning another feature of this large-$N$ critical
behavior which emerges from a numerical analysis of the phase
distribution of the eigenvalues of the link operator
\begin{equation}
L = U_x \, U^\dagger_{x+\mu}:
\end{equation}
the $N=\infty$ phase transition should be related to the
compactification of the eigenvalues of $L$
\cite{Campostrini-Rossi-Vicari-chiral-3}, like the Gross-Witten phase
transition.

The existence of such a phase transition does not represent an
obstruction to the use of strong-coupling expansion for the
investigation of the continuum limit.  Indeed large-$N$ Monte Carlo
data show scaling and asymptotic scaling (in the energy scheme) even
for $\beta$ smaller then the peak of the specific heat, suggesting an
effective decoupling of the modes responsible for the large-$N$ phase
transition from those determining the physical continuum limit.  This
fact opens the road to tests of scaling and asymptotic scaling at
$N=\infty$ based only on strong-coupling computations, given that the
strong-coupling expansion should converge for $\beta < \beta_c$.  (The
strong-coupling analysis does not show evidence of singularities in
the complex $\beta$-plane closer to the origin than $\beta_c$.)

In the continuum limit the dimensionless renormalization-group
invariant function
\begin{equation}
A(p;\beta)\equiv{\widetilde{G}(0;\beta)\over\widetilde{G}(p;\beta)}
\label{ldef}
\end{equation}
turns into a function $A(y)$ of the ratio $y\equiv p^2/M_G^2$ only,
where $M_G^2\equiv 1/\xi_G^2$ and $\xi_G$ is the second moment
correlation length
\begin{equation}
\xi_G^2\equiv {1\over 4}\,{\sum_x x^2G(x)\over \sum_x G(x)}.
\label{xig}
\end{equation}
$A(y)$ can be expanded in powers of $y$ around $y=0$:
\begin{equation}
A(y)=1 + y + \sum_{i=2}^\infty c_i y^i,
\label{lexp}
\end{equation}
and the coefficients $c_i$ parameterize the difference from a
generalized Gaussian propagator.  The zero $y_0$ of $A(y)$ closest to
the origin is related to the ratio $M^2/M_G^2$, where $M$ is the
fundamental mass; indeed $y_0=-M^2/M_G^2$.  $M^2/M_G^2$ is in general
different from one; it is one in Gaussian models (i.e. when
$A(y)=1 + y $).

Numerical simulations at large $N$, which allow an investigation of
the region $y\geq 0$, have shown that the large-$N$ limit of the
function $A(y)$ is approached rapidly and that its behavior is
essentially Gaussian for $y \lesssim 1$, indicating that $c_i\ll 1$ in
Eq.\ (\ref{lexp}) \cite{Rossi-Vicari-chiral1}.  Important logarithmic
corrections to the Gaussian behavior must eventually appear at
sufficiently large momenta, as predicted by simple weak-coupling
calculations supplemented by a renormalization group resummation:
\begin{equation}
\widetilde{G}(p)\sim {\log p^2 \over p^2} 
\label{gpert}
\end{equation}
for $p^2/M_G^2\gg 1$ and in the large-$N$ limit.

The approximate Gaussian behavior at small momentum is also confirmed
by the direct estimate of the ratio $M^2/M_G^2$ obtained by
extrapolating Monte Carlo data to $N=\infty$.  The large-$N$ limit of
the ratio $M^2/M_G^2$ is rapidly approached, already at $N=6$ within
few per mille, leading to the estimate $M^2/M_G^2=0.982(2)$, which is
very close to one \cite{Rossi-Vicari-chiral2}.  Large-$N$
strong-coupling computations of $M^2/M_G^2$ provide a quite stable
curve for a large region of values of the correlation length, which
agrees (within about one per cent) with the continuum large-$N$ value
extrapolated by Monte Carlo data
\cite{Campostrini-Rossi-Vicari-chiral-2}.

Monte Carlo simulations at large values of $N$ ($N\geq 6$) also show
that asymptotic scaling predictions applied to the fundamental mass
are verified within a few per cent at relatively small values of the
correlation length ($\xi\gtrsim2$) and even before the peak of the
specific heat in the so-called ``energy scheme'' \cite{Parisi-betaE};
the energy scheme is obtained by replacing $T$ with a new temperature
variable $T_E \propto E$, where $E$ is the internal energy density.
At $N=\infty$ a test of asymptotic scaling may be performed by using
the large-$N$ strong-coupling series of the fundamental mass.  The
two-loop renormalization group and a Bethe Ansatz evaluation of the
mass/$\Lambda$-parameter ratio \cite{Balog-Naik-Niedermayer-Weisz}
lead to the following large-$N$ asymptotic scaling prediction in the
$\beta_E$ scheme:
\begin{eqnarray}
&&M \cong 16\,\sqrt{\pi\over\e} \exp\!\left(\pi\over4\right)
\Lambda_{E,2l}(\beta_E),\nonumber \\ 
&&\Lambda_{E,2l}(\beta_E)  = 
\sqrt{8\pi\beta_E}\exp(-8\pi\beta_E) ,\nonumber \\
&&\beta_E = {1\over 8E}\,.
\label{mass-lambdaE}
\end{eqnarray}
Strong-coupling calculations, where the new coupling $\beta_E$ is
extracted from the strong-coupling series of $E$, show asymptotic
scaling within about 5\% in a relatively large region of values of the
correlation length ($1.5\lesssim\xi\lesssim3$)
\cite{Campostrini-Rossi-Vicari-chiral-brief,%
Campostrini-Rossi-Vicari-chiral-2}.

The good behavior of the large-$N$ $\beta$-function in the $\beta_E$
scheme, and therefore the fact that physical quantities appear to be
smooth functions of the energy, together with the critical behavior
(\ref{Ccrit}), can be explained by the existence of a non-analytical
zero at $\beta_c$ of the $\beta$-function in the standard scheme:
\begin{equation}
\beta_L(T)\equiv a{\d T\over\d a}\sim |\beta-\beta_c|^\alpha
\label{betasing}
\end{equation}
around $\beta_c$, where $\alpha$ is the critical exponent of the
specific heat.  This is also confirmed by an analysis of the
strong-coupling series of the magnetic susceptibility $\chi$ and
$M^2_G$, which supports the relations
\begin{equation}
{\d\log\chi\over\d\beta} \sim  
{\d\log M^2_G\over\d\beta} \sim  
|\beta-\beta_c|^{-\alpha}
\label{chi_crit}
\end{equation}
in the neighborhood of $\beta_c$, which are consequences of
Eq.\ (\ref{betasing}) \cite{Campostrini-Rossi-Vicari-chiral-2}.

We finally mention that similar results have been obtained for
two-dimensional chiral models on the honeycomb lattice by a large-$N$
strong-coupling analysis. In fact an analysis of the 26th-order
strong-coupling series of the free energy indicates the presence of a
large-$N$ phase transition, with specific heat exponent $\alpha \cong
0.17$, not far from that found on the square lattice (we have no
reasons to expect that the large-$N$ phase transition on the square
and honeycomb lattices are in the same universality class).
Furthermore the mass-gap extracted from the 20th-order strong-coupling
expansion of $G(x)$ allows to check the corresponding asymptotic
scaling predictions in the energy scheme within about 10\%
\cite{Campostrini-Rossi-Vicari-chiral-2}.

\subsection{The large-$N$ limit of ${\rm SU}(N)$
  lattice gauge theories}
\label{secQCD}

An overview of the large-$N$ limit of the continuum formulation of QCD
has been already presented in Sect.\ \ref{unitary-matrices}.  In the
following we report some results concerning the lattice approach.

Gauge models on the lattice have been mostly studied in their Wilson
formulation
\begin{eqnarray}
S_{\rm W} &=& N\beta \sum_{x,\mu>\nu} {\rm Tr} \left[
U_\mu(x) U_\nu(x+\mu) U_\mu^\dagger(x+\nu) U_\nu^\dagger(x) 
+ {\rm h.c.}\right]. \nonumber \\
\label{wilsonac}
\end{eqnarray}
In view of a large-$N$ analysis one may consider both ${\rm SU}(N)$
and ${\rm U}(N)$ models, since they are expected to reproduce the same
statistical theory in the limit $N\rightarrow\infty$ (at fixed
$\beta$).  As for two-dimensional chiral models, ${\rm SU}(N)$ and
${\rm U}(N)$ models should have the same continuum limit for any
finite $N\geq 2$.

The phase diagram of statistical models defined by the Wilson action
has been investigated by standard techniques, i.e., strong-coupling
expansion, mean field \cite{Drouffe-Zuber}, and Monte Carlo
simulations
\cite{Creutz-SU5,Creutz-Moriarty,Moriarty-Samuel-comparison}. These
studies show the presence of a first-order phase transition in ${\rm
SU}(N)$ models for $N\geq 4$, and in ${\rm U}(N)$ models for any
finite $N$.  A first-order phase transition is then expected also in
the large-$N$ limit at a finite value of $\beta$, which is estimated
to be $\beta_c \approx 0.38$ by mean-field calculations and by
extrapolation of Monte Carlo results.  A review of these results can
be found in Ref.\ \cite{Itzykson-Drouffe}.  Some speculations on the
large-$N$ phase diagram can be also found in Refs.\ 
\cite{Kostov-sc,Green-Samuel-largeN}.  The r\^ole of the determinant
of Wilson loops in the phase transition of ${\rm U}(N)$ gauge models
has been investigated in Ref.\ \cite{Green-Samuel-largeN} by
strong-coupling character expansion, and in Ref.\ 
\cite{Moriarty-Samuel} by Monte Carlo simulations.

Large-$N$ mean-field calculations suggest the persistence of a
first-order phase transition when an adjoint-representation coupling
is added to the Wilson action
\cite{Chen-Tan-Zheng-phase,Ogilvie-Horowitz}.

The first-order phase transition of ${\rm SU}(N)$ lattice models at
$N>3$ can probably be avoided by choosing appropriate lattice actions
closer to the renormalization group trajectory of the continuum limit,
as shown in Ref.\ \cite{Itoh-Iwasaki-Yoshie-absence} for ${\rm
SU}(5)$.  In ${\rm U}(N)$ models the use of such improved actions
should leave a residual transition, due to the extra ${\rm U}(1)$
degrees of freedom which should decouple at large $\beta$ in order to
reproduce the physical continuum limit of ${\rm SU}(N)$ gauge models.

It is worth mentioning two studies of confinement properties at large
$N$, obtained essentially by strong-coupling arguments.  In Ref.\ 
\cite{Greensite-Halpern}, the authors argue that deconfinement of
heavy adjoint quarks by color screening is suppressed in the large-$N$
limit.  At $N=\infty$, the adjoint string tension is expected to be
twice the fundamental string tension, as implied by factorization.
In Ref.\ \cite{Lovelace-universality}, strong-coupling based arguments
point out that Wilson loops in ${\rm O}(N)$, ${\rm U}(N)$, and ${\rm
Sp}(N)$ lattice gauge theories should have the same large-$N$ limit,
and therefore these theories should share the same confinement
mechanism.  Such results should be taken into account when studying
confinement mechanisms.

Studies based on Monte Carlo simulations for $N>3$ have not gone
beyond an investigation of the phase diagram, so no results concerning
the continuum limit of ${\rm SU}(N)$ lattice gauge theories with $N>3$
have been produced.  Estimates of the mass of the lightest glueball,
obtained by a variational approach within a Hamiltonian lattice
formulation, seem to indicate a rapid convergence of the $1/N$
expansion \cite{Chin-Karliner}.

An important breakthrough for the study of the large-$N$ limit of
${\rm SU}(N)$ gauge theories has been the introduction of the
so-called reduced models. A quite complete review on this subject can
be found in Ref.\ \cite{Das-review}.  

Eguchi and Kawai \cite{Eguchi-Kawai-reduction} pointed out that, as a
consequence of the large-$N$ factorization, one can construct
one-site theories equivalent to lattice YM in the limit
$N\rightarrow\infty$.  The simplest example is given by the one-site
matrix model obtained by replacing all link variables of the standard
Wilson formulation with four ${\rm SU}(N)$ matrices according to the
simple rule
\begin{equation}
U_\mu(x) \to U_\mu.
\label{EK}
\end{equation}
This leads to the reduced action
\begin{equation}
S_{\rm EK}= N\beta \sum_{\mu>\nu}{\rm Tr} \left[
U_\mu U_\nu U_\mu^\dagger U_\nu^\dagger + {\rm h.c.}\right].
\label{EKaction}
\end{equation}
Reduced operators, and in particular reduced Wilson loops, can be
constructed using the correspondence (\ref{EK}).  In the large-$N$
limit one can prove that expectation values of reduced Wilson loop
operators satisfy the same Schwinger-Dyson equations as those in the
Wilson formulation.  Assuming that all features of the $N=\infty$
theory are captured by the Schwinger-Dyson equations of Wilson loops,
the reduced model may provide a model equivalent to the standard
Wilson theory at $N=\infty$.  In the proof of this equivalence the
residual symmetry of the reduced model
\begin{equation}
U_\mu \to Z_\mu U_\mu, \qquad Z_\mu\in Z_N,
\label{Zn}
\end{equation}
where $Z_N$ is the center of the ${\rm SU}(N)$ group, plays a crucial
r\^ole.  Therefore, the equivalence in the large-$N$ limit of the
Wilson formulation and the reduced model (\ref{EKaction}) is actually
valid if the symmetry (\ref{Zn}) is unbroken. This is verified only in
the strong-coupling region; indeed in the weak-coupling region the
$Z_N^4$ symmetry gets spontaneously broken and therefore the
equivalence cannot be extended to weak coupling
\cite{Bhanot-Heller-Neuberger}.

In order to avoid this unwanted phenomenon of symmetry breaking and to
extend the equivalence to the most interesting region of the continuum
limit, modifications of the original Eguchi-Kawai model have been
proposed \cite{Eguchi-Nakayama-simplification,Bhanot-Heller-Neuberger,%
GonzalezArroyo-Okawa-twisted}.  The most promising one for numerical
simulation is the so-called twisted Eguchi-Kawai (TEK)
model \cite{Eguchi-Nakayama-simplification,%
GonzalezArroyo-Okawa-twisted}.  Instead of the correspondence
(\ref{EK}), the twisted reduction prescription consists in replacing
\begin{equation}
U_\mu(x) \to T(x)U_\mu T(x)^\dagger,
\label{TEK}
\end{equation}
where
\begin{equation}
T(x)= \prod_\mu (\Gamma_\mu)^{x_\mu}
\label{trasl}
\end{equation}
and $\Gamma_\mu$ are traceless ${\rm SU}(N)$ matrices obeying the
't Hooft algebra
\begin{equation}
\Gamma_\nu \Gamma_\mu = Z_{\mu\nu}\Gamma_\mu \Gamma_\nu ;
\label{twist}
\end{equation}
$Z_{\mu\nu}$ is an element of the center of the group $Z_N$,
\begin{equation}
Z_{\mu\nu} = \exp\!\left(\I{2\pi\over N} n_{\mu\nu}\right),
\end{equation}
where $n_{\mu\nu}$ is an antisymmetric tensor with $n_{\mu\nu}=1$ for
$\mu<\nu$.  $\Gamma_\mu$ are the matrices implementing the
translations by one lattice spacing in the $\mu$ direction (here it is
crucial that the fields $U_\mu$ are in the adjoint representation).
The twisted reduction applied to the Wilson action leads to the
reduced action
\begin{equation}
S_{\rm TEK}= N\beta \sum_{\mu>\nu}
{\rm Tr}\left[ Z_{\mu\nu}U_\mu U_\nu U_\mu^\dagger U_\nu^\dagger 
+ {\rm h.c.}\right].
\label{TEKaction}
\end{equation}
The correspondence between correlation functions of the large-$N$ pure
gauge theory and those of the reduced twisted model is obtained as
follows.  Let ${\cal A}[U_\mu(x)]$ be any gauge invariant functional
of the field $U_\mu(x)$, then
\begin{equation}
\langle {\cal A}[U_\mu(x)]\rangle_{[N=\infty,\ {\rm YM}]}=
\langle {\cal A}[ T(x)U_\mu T(x)^\dagger]
\rangle_{[N=\infty,\ {\rm TEK}]}
\label{TEKfunc}
\end{equation}
Once again the Schwinger-Dyson equations for the reduced Wilson loops,
constructed using the correspondence (\ref{TEK}), are identical to the
loop equations in the Wilson formulation when $N\rightarrow\infty$.
The residual symmetry (\ref{Zn}), which is again crucial in the proof
of the equivalence, should not be broken in the weak-coupling region,
and therefore the equivalence should be complete in this case.

One can also show that:

(i) the reduced TEK model is equivalent to the corresponding field
theory on a periodic box of size $L=\sqrt{N}$ \cite{Das-review};

(ii) in the large-$N$ limit finite-$N$ corrections are $O(1/N^2)$,
just as in the ${\rm SU}(N)$ lattice gauge theory.

Moreover, since $N^2=L^4$, finite-$N$ corrections can be seen as
finite-volume corrections. Therefore in twisted reduced models the
large-$N$ and thermodynamic limits are connected and approached
simultaneously.

Monte Carlo studies of twisted reduced models at large $N$ confirm the
existence of a first-order phase transition at $N=\infty$ located at
$\beta_c = 0.36(2)$ \cite{GonzalezArroyo-Okawa-string}, which is
consistent with the mean-field prediction $\beta_c\simeq 0.38$
\cite{Itzykson-Drouffe}.  This transition is a bulk transition, and it
does not spoil confinement.  The few and relatively old existing Monte
Carlo results obtained in the weak-coupling region (cfr.\ e.g.\ Refs.\ 
\cite{GonzalezArroyo-Okawa-string,Fabricius-Haan,Haan-Meier}) seem to
support a rapid approach to the $N\rightarrow\infty$ limit of the
physical quantities, and are relatively close to the corresponding
results for ${\rm SU}(3)$ obtained by performing simulations within
the Wilson formulation. This would indicate that $N=3$ is sufficiently
large to consider the large-$N$ limit a good approximation of the
theory.

We mention that hot twisted models can be constructed, which should be
equivalent to QCD at finite temperature in the large-$N$ limit (cfr.\  
Ref.\ \cite{Das-review} for details on this subject).



\end{document}